\documentclass[prd,aps,12pt]{revtex4}
\usepackage{epsfig}
\begin{document}

\title{%
\hfill{\normalsize\vbox{%
\hbox{\rm May 2003}
\hbox{\rm JLAB-THY-03-32}
\hbox{\rm SU-4252-778} }}\\
\vspace{0.5cm}
{\bf Comparing the Higgs sector of electroweak theory with \\ the scalar sector of low energy QCD.}}
\author{{\bf Abdou Abdel-Rehim}$^{\rm  (a)}$}
\email{abdou@physics.syr.edu} 

\author{{\bf Deirdre Black}$^{\rm  (b)}$}
\email{dblack@jlab.org}
 
\author{{\bf Amir H. Fariborz}$^{\rm  (c)}$}
\email{fariboa@sunyit.edu}

\author{{\bf Salah Nasri}$^{\rm  (a,d)}$}
\email{snasri@physics.syr.edu}

\author{{\bf Joseph Schechter}$^{\rm  (a)}$}
\email{schechte@physics.syr.edu}

\affiliation{$^{\rm (a)}$ Department of Physics, Syracuse University,
Syracuse, NY 13244-1130}
\affiliation{$^{\rm (b)}$ Jefferson Lab, 12000 Jefferson
Ave., Newport News, VA 23606. }
\affiliation{$^{\rm (c)}$ Department of Mathematics/Science, State
University of New York Institute of Technology, Utica, NY 13504-3050}
\affiliation{$^{\rm (d)}$ Kavli Institute for Theoretical
Physics, University of California, Santa Barbara, CA 93106}

\begin{abstract}
 We first review how the simple K-matrix unitarized linear
SU(2) sigma model can explain the experimental
data in the scalar $\pi \pi$ scattering channel of QCD up to about 800 MeV. Since it is 
just a scaled version of the minimal electroweak Higgs sector, which is
often treated with the same unitarization method, we 
interpret the result as
 support for this approach in the electroweak model 
with scaled values of tree level Higgs mass up to at least about 2 TeV.
 We further note that the   
 relevant QCD effective Lagrangian which fits the data
to still higher energies
using the same method involves another scalar resonance. This suggests 
that the
method should also be applicable to corresponding beyond minimal
electroweak models.
Nevertheless we note that even with one resonance, the minimal
K matrix unitarized model behaves smoothly for large bare Higgs mass  
by effectively "integrating out" the Higgs while preserving unitarity.
 With added confidence in this simple approach we make a survey of
the Higgs sector for the full range of bare Higgs mass. 
One amusing point which emerges is that the 
characteristic factor of the W-W fusion mechanism for
Higgs production peaks at the bare mass of the Higgs boson,
while the characteristic factor for the gluon fusion mechanism
peaks near the generally lower {\it physical} mass.

\end{abstract}

\maketitle

\section{Introduction}

There has recently been renewed interest \cite{kyotoconf}-
\cite{BFMNS01}
 in the low energy scalar
sector of QCD. Especially, many physicists now believe in the
existence of the light, broad $I=J=0$ resonance, sigma in the
500-600 MeV region. The sigma is also very likely the "tip of an 
iceberg" which may consist of a nonet of light non $q{\bar q}$ -
type scalars \cite{Jaffe}
 mixing \cite{mixing}
 with another heavier $q \bar q$ - type
scalar nonet as well as a glueball.

A variety of different approaches to meson-meson scattering have
been employed to argue for a picture of this sort. For our present 
purpose, first consider an approach to describing $\pi\pi$, $I=J=0$
scattering which is based on a conventional non-linear chiral 
Lagrangian of pseudoscalar and vector fields augmented by scalar
fields also introduced in a chiral invariant manner \cite{HSS1}.
 The experimental
data from threshold to a bit beyond 1 GeV can be fit by computing the
tree amplitude from this Lagrangian and making an approximate
unitarization. The following ingredients are present: i) the
"current algebra" contact term, ii) the vector meson exchange terms,
iii) the unitarized $\sigma$(560) exchange terms and iv) the unitarized
$f_0$(980) exchange terms. Although the $\rho$(770) vector
meson certainly is a crucial feature of low energy physics,
it is amusing to note that a fit can be made \cite{HSS2}
 in which the 
contribution ii) is absent. This results in a somewhat lighter and
broader sigma meson, in agreement with other approaches which neglect the 
effect of the $\rho$ meson.

For the purpose of checking this strong interaction calculation, the
meson-meson scattering was also calculated in a general version of
the {\it linear} SU(3) sigma model, which contains both $\sigma$(560)
 and $f_0(980)$ candidates \cite{BFMNS01}.
 The procedure adopted was to calculate the
tree amplitude and then to unitarize, without introducing any new 
parameters, by identifying the tree amplitude as the K-matrix
amplitude. This also gave a reasonable fit to the data, including the
characteristic Ramsauer-Townsend effect which flips the sign of the
$f_0(980)$ resonance contribution. At a deeper and more realistic level of
description in the linear sigma model framework, one expects two different
chiral multiplets - $q{\bar q}$ as well as $qq{\bar q}{\bar q}$ - 
to mix with each other. A start on this model seems encouraging
\cite{sectionV}.

Now, if we restrict attention to the energy range from threshold to about
800 MeV [before the $f_0(980)$ becomes important] the data are well fit
by the simple SU(2) linear sigma model \cite{GL},
 when unitarized by the
K-matrix method \cite{AS94}.
 If one further restricts attention to the energy range
from $\pi\pi$ threshold to about 450 MeV, the data can be fit by using
the non-linear SU(2) sigma model \cite{nonlinear},
 which contains only pion fields at tree
or one loop (chiral perturbation theory) level. To get a description 
of the sigma resonance region by using only chiral perturbation theory 
\cite{chpt}
would seem to require a prohibitively large number of loops
(see for example \cite{loops}).

The lessons we draw are, first, that the {\it prescription} of using the 
K-matrix unitarized SU(2) linear sigma model provides one with a
simple explanation of the scalar sector of QCD in its non-perturbative
low energy region. Secondly, the sigma particle of this model is not
necessarily just a $ q {\bar q}$ bound state in the underlying fundamental
QCD theory. Rather the linear SU(2) sigma model seems to be a "robust"
framework for describing the spontaneous breakdown of $SU(2)_L \times 
SU(2)_R \sim O(4)$ to $SU(2)$. There are just two parameters.
One may be taken to be the vacuum value of the sigma field; this is
proportional to the "pion decay constant" and sets the low energy
scale for the theory. The second may be taken to be the "bare"
mass of the sigma particle; as its value increases the theory becomes
more strongly interacting and hence non-perturbative. The K-matrix
unitarization gives a physically sensible prescription for this
non-perturbative regime wherein the model predicts the "physical"
mass of the sigma to be significantly smaller than the "bare" mass. 
There is in fact a maximum predicted "physical" mass as the "bare" mass
is varied.

Now it is well known that the Higgs sector of the standard electroweak model
is identical to the SU(2) linear sigma model of mesons. The sigma corresponds
to the Higgs boson while the $\pi^{\pm}$ and $\pi^0$ appear
eventually, in the Unitary gauge, as the longitudinal components of the $W^{\pm}$ and Z bosons. There is
an important difference of scale in that the vacuum expectation value of the
Higgs boson field is about 2656 times that of the "QCD sigma". However,
once the bare masses of the two theories are scaled to their
corresponding VEV's, one can formally treat both applications of the
same model at once. The practical significance of comparing
these two applications is enhanced by the Goldstone boson equivalence
theorem \cite{CLT74}-\cite{CDP02}.
 This theorem has the implication that, for energies where
$M_W/E$ is small, the important physical amplitudes involving longitudinal 
W and Z bosons are given by the corresponding amplitudes of the 
sub-Lagrangian of the electroweak theory obtained by deleting everything
except the scalar fields (massless "pions" and massive Higgs boson).
Thus, except for rescaling, the electroweak amplitudes may be directly 
given by the pion amplitudes computed for QCD with varying bare sigma 
mass. There is no reason why the scaled Higgs boson mass should
be the same as the scaled "QCD sigma" mass since, while the two 
Lagrangians agree, the scaled bare mass is a free parameter.

A simple and perhaps most important point of the present paper
is that the same model which appears to effectively describe the
Higgs sector of the electroweak model can be used, with the K matrix 
prescription  to describe the non-perturbative scalar sector of
QCD {\it in agreement with experiment}. In fact, the K-matrix method of
unitarization is a popular \cite{RS89}-\cite{W91} approach to the non-
perturbative regime of the electroweak model, although it is not
easy to rigorously justify. We will see that the QCD sigma meson
has about the same scaled mass as a Higgs boson of 2 TeV. This
provides a practical justification of the use of the K-matrix
procedure up to that value at least. In fact the successful
addition of the $f_0(980)$ resonance to the low energy sigma
using the linear SU(3) sigma model and the same unitarization scheme
suggests that the range of validity of the electroweak treatment is even 
higher. It also suggests that the same method should also work in
models which have more than one Higgs meson in the same channel.
Furthermore, we will note that, treating the unitarized model as a 
"prescription", even the bare mass going to infinity is a sensible limit.

A leading experimental question at the moment concerns the actual
mass of the Higgs particle. Direct experimental search \cite{C02}
rules out values less than about 115 GeV. Indirect observation
via a "precision" global analysis of all electroweak data including
the effects of virtual Higgs boson exchange in loop diagrams actually
gives a preferred central value of about 80 GeV \cite{LEP}. However there
are unexplained "precision" effects like the NuTeV experiment
on $\mu$ neutrino scattering off nucleons \cite{NuT} and the 
measurement of
the invisible Z width at LEP/SLC \cite{LEP} which raise doubts about 
what is happening. For example in ref \cite{LOTW}, the
authors suggest the possibility of explaining these two effects by 
reducing the strength of neutrino couplings to the Z boson. In
that case, they find a new overall fit which prefers rather large Higgs 
masses, even up to about 1 TeV. This is somewhat speculative but
does make it timely to revisit the large mass Higgs sector 
and the consequent need for unitarity corrections.

In the present paper, after giving some notation in section II,
we review in section III the fit to low energy $I=J=0$ $\pi\pi$
scattering obtained with the K-matrix unitarized SU(2)
linear sigma model amplitude and its extension to higher energies 
using an SU(3) linear sigma model. This establishes a preferred 
value for the bare mass of the "QCD sigma" and shows it to
be in the non-perturbative region of the linear sigma model.
The physical mass and width, obtained from the pole position in
the complex s plane, differ appreciably from their "bare" or tree
level values. This effect is explored in section IV for a full range of
sigma "bare" mass values, which can be rescaled to arbitrary choices
of bare Higgs mass in the electroweak theory. The difference between
zero and non-zero pion mass is also shown to be qualitatively small.
Further, the existence of a maximum value for the physical sigma mass
is displayed.

In section V, using the equivalence theorem, we discuss the scattering
in the $I=J=0$ (Higgs) channel of longitudinal gauge bosons for the
complete range of bare Higgs masses. This provides the characteristic
factor in the proposed "W-fusion" mechanism \cite{DGMP97}
 for Higgs production,
although here we will mainly regard it as a "thought experiment". Some
of this material has already been given by other workers
 \cite{RS89}-\cite{W91}.
One feature which is perhaps treated in more detail here is the
peculiar behavior of the scattering amplitude for large bare Higgs
mass values. We note that it can be thought of as an evolution
to the characteristic bare mass $=$ infinity shape. This shape is shown 
to be simply the K-matrix unitarized amplitude of the {\it non linear
sigma model}. In other words, taking the bare mass to infinity in the 
K-matrix unitarized amplitude effectively "integrates out" the
sigma field. We also show that the magnitude of the scattering amplitude 
always peaks at the {\it bare} Higgs mass. For values greater than about 3 TeV,
the width is so great that the peaking is essentially unobservable. 

In section VI we discuss the characteristic factor for the "gluon fusion"
reaction \cite{GGMN78},
 which is another possible source of Higgs boson production.
Unlike the W-fusion reaction, this involves final state interactions
rather than unitarization. We point out that in our model the magnitude
of the gluon-fusion factor peaks at the {\it physical} Higgs mass
rather than at the bare Higgs mass found in the W-fusion case. It
seems to present an amusing example showing how the production mechanism
of the Higgs boson can influence its perceived properties. Finally some
concluding remarks and directions for future work are offered.

\section{Comparison of notations}
First let us make a correspondence between the notations employed for
the ${\rm SU}(2)_L \times {\rm SU}(2)_R$ linear sigma model used to
model low energy QCD and to model the Higgs sector of the 
minimal standard
${\rm SU}(2)_L \times {\rm U}(1)$ electroweak theory.  In the latter
case the Higgs sector by itself possesses $O(4)
\sim {\rm SU}(2)_L \times {\rm SU}(2)_R$ symmetry which is explicitly
broken when the gauge bosons and fermions are taken into account.  In
the low energy QCD application the Lagrangian density is written in
terms of the pion and sigma as
\begin{equation}
{\cal{L}} = -\frac{1}{2} \left( \partial_\mu {\mbox{\boldmath ${\pi}$
}} \cdot \partial_\mu {\mbox{\boldmath ${\pi}$}} + \partial_\mu \sigma
\partial_\mu \sigma \right) + a {\left
( \sigma^2 +  {\mbox{\boldmath ${\pi}$ }}^2 \right)} - b  {\left
( \sigma^2 +  {\mbox{\boldmath ${\pi}$ }}^2 \right)}^2 ,
\label{LsMLag}
\end{equation}
where the real parameters $a$ and $b$ are both taken positive to
insure spontaneous breakdown of chiral symmetry.  The vacuum value
$\left< \sigma \right>$ of the $\sigma$ field is related to the pion
decay constant as 
\begin{equation}
{\rm F}_\pi = \sqrt{2} \left< \sigma \right>.
\label{Fpi}
\end{equation}
(In the low energy QCD model, ${\rm F}_\pi = 0.131$ GeV). 
The parameter $a$ is given by 
\begin{equation}
a = \frac{1}{4} m_{\sigma b}^2,
\end{equation}
where $m_{\sigma b}$ is the tree level value of the sigma mass.
Finally the dimensionless parameter $b$, whose value furnishes a
criterion for the applicability of perturbation theory, is related to
other quantities as
\begin{equation}
b = \frac{m_{\sigma b}^2}{8 {\left< \sigma \right>}^2}.
\label{b}
\end{equation}

The Higgs sector of the standard model employs the fields
\begin{equation}
\Phi = \left( \begin{array}{c} \phi^+ \\ \phi^0 \end{array} \right)
\quad , \quad \Phi^\dagger = \left( \begin{array}{c c} \phi^- &
{\phi^0}^* \end{array} \right).
\end{equation}
These can be rewritten in terms of the $\mbox{\boldmath $\pi$}$ and
$\sigma$ fields by identifying
\begin{equation}
\Phi = \left( \begin{array}{c} i \pi^+ \\ \frac{\sigma - i
\pi^0}{\sqrt 2} \end{array} \right).
\label{identification}
\end{equation}
In this language the same Lagrangian, Eq. (\ref{LsMLag}) is written 
\begin{equation}
{\cal{L}} = - \partial_\mu \Phi^\dagger \partial_\mu \Phi + 2a
\Phi^\dagger \Phi - 4b { (\Phi^\dagger \Phi)}^2.
\label{HiggsLag}
\end{equation}
Here one employs the notation $v= \sqrt{2} \left< \phi^0 \right>$ so that, noticing
Eq. (\ref{identification})
\begin{equation}
v = \left< \sigma \right> = \frac{F_\pi}{\sqrt 2}.
\end{equation}
(In the electroweak theory, $v = 0.246$ TeV, about 2656 times the value
in the low energy QCD case).  

\section{Unitarized linear sigma model}

Now let us consider the SU(2) linear sigma model in
Eq. (\ref{LsMLag}) as a ``toy model'' for the scattering of two pions
in the s-wave iso-singlet channel.  It has been known for many
years that this gives a good description of the low energy scattering
near threshold in the limit of large sigma mass.  Of course, the
 non-linear sigma model is more convenient for chiral
perturbation theory calculations.  However we are going to focus on
the properties of the sigma here and the simple linear sigma model is
appropriate for this purpose.  

The I=J=0 partial wave amplitude at tree level is 

\begin{equation}
\left[T^{0}_{0}\right]_{\rm tree}(s) =  \alpha \left( {s}
\right) + \frac{\beta (s)}{m_{\sigma b}^2 - s}
\label{pipiampl}
\end{equation}
where 
\begin{eqnarray}
\alpha \left( \rm{s}\right) &=&  \frac { \sqrt {1 -
\frac{4m_\pi^2}{s}}}{32\pi F_\pi^2} \left({m_{\sigma b}^2 } -
{m_\pi}^2 \right)
\left[ -10 + 4 \frac{{m_{\sigma b}^2} - {m_\pi}^2}{s - 4 {m_\pi}^2} \rm{ln}
\left( \frac{{m_{\sigma b}^2} + s - 4{m_\pi}^2}{{m_{\sigma b}^2}} \right) \right],  \nonumber \\
\beta (s) &=& \frac {3 \sqrt {1 -
\frac{4m_\pi^2}{s}}}{16 \pi F_\pi^2}   {\left( {m_{\sigma b}^2} - {m_\pi}^2 \right)}^2.
\label{alphabeta}
\end{eqnarray}
For generality we have added the effect of a non-zero pion mass which
would correspond to the addition of a small term linear in $\sigma$ to
Eq. (\ref{LsMLag}).  The normalization of the amplitude $T_0^0(s)$ is
given by its relation to the partial wave S-matrix
\begin{equation}
S_0^0(s) = 1 + 2iT_0^0(s).
\label{Smatrix}
\end{equation}
The formula above is not, of course, new; we are following
here the notations of \cite{BFMNS01}, which contains many
references.  While this tree-level formula
works well at threshold it does involve large coupling constants and
cannot be expected to be a priori reasonable even several hundred MeV
above threshold.  In addition, at the point $s=m_{\sigma b}^2$, the
amplitude Eq. (\ref{pipiampl}) diverges.  A usual solution to this
problem is to include a phenomenological width term in the denominator
by making the replacement: 
\begin{equation}
\frac{1} {m_{\sigma b}^2  - s} \longrightarrow \frac{1}
{m_{\sigma b}^2  - s - i m_{\sigma b}\Gamma }.
\label{conventionalreg}
\end{equation}
However this standard approach is not a good idea in the present
case.  As emphasized especially by Achasov and Shestakov \cite{AS94},
the replacement Eq. (\ref{conventionalreg}) completely destroys the
good threshold result.  This is readily understandable since the
threshold result is well known to arise from a nearly complete
cancellation between the first and second terms of
Eq. (\ref{pipiampl}).  An advantage of the non-linear sigma model is
that the good threshold result is obtained directly without need for
such a delicate cancellation.  However, the pole in the linear sigma
model can be successfully handled by using, instead of
Eq. (\ref{conventionalreg}), K-matrix
 regularization \cite{AS94,Chung}, which instructs
us to adopt the manifestly unitary form   
\begin{equation}
S_0^0 (s) = \frac {1 + i \left[{T^{0}_{0}}\right]_{\rm tree}(s) }{1 -
i \left[{T^{0}_{0}}\right]_{\rm tree} (s) } 
\label{regularization}
\end{equation}
Using Eq. (\ref{Smatrix}) we get
\begin{equation}
T_0^0(s) = \frac { \left[{T^{0}_{0}}\right]_{tree}(s)}{ 1 - i
\left[{T^{0}_{0}}\right]_{tree}(s)}.
\label{Tregularization}
\end{equation}
Near threshold, where $\left[{T^{0}_{0}(s)}\right]_{tree}$ is small,
this reduces to $\left[{T^{0}_{0}(s)}\right]_{tree}$ as desired.
Elsewhere it provides a unitarization of the theory which is seen to
have the general structure of a ``bubble-sum''.  We will adopt this
very simple model as a provisional approximation for the strong
coupling regime of QCD in the I=J=0 channel. 

\begin{figure}[htbp]
\vspace{0.5cm}
\begin{center}
\mbox{\epsfig{file=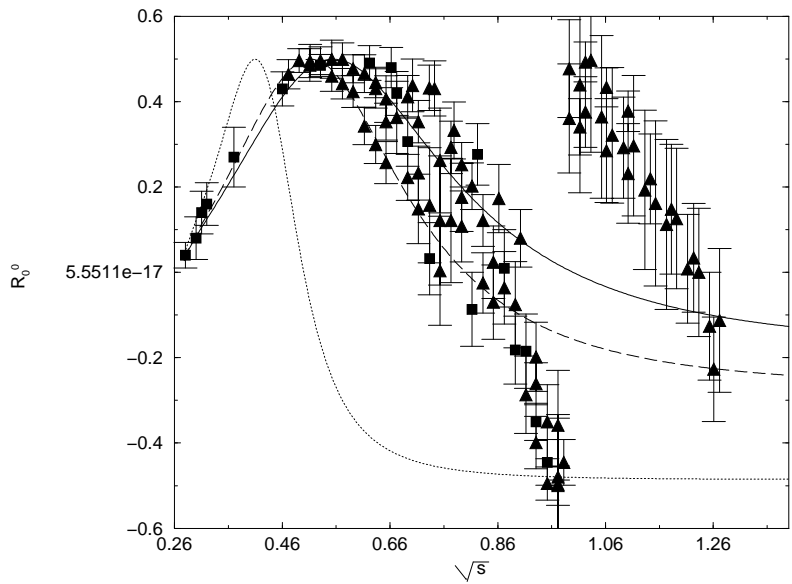, height=10cm, angle=270}}
\end{center}
\caption[]
{Comparison with experiment of Real part of the I=J=0 $\pi \pi$ scattering
amplitude in the SU(2) Linear Sigma Model, for $m_{\sigma b} =0.5$ GeV (dots),
$m_{\sigma b} =0.8$ GeV (dashes) and $m_{\sigma b}=1$
GeV (solid).  Figure taken from \cite{BFMNS01}.  Experimental data are extracted from Alekseeva
{\it et al} (squares) and Grayer {\it et al} (triangles) \cite{pipidata}.}
\label{finalSU2LsMRamp}
\end{figure}

It seems difficult to rigorously justify this as an effective procedure for the
strong coupling regime.  However we may at least compare with the
experimental data on $\pi \pi$ scattering \cite{pipidata}.
  Since $F_\pi$ is known there
is only a single parameter - the tree-level mass $m_{\sigma b}$.  In
Fig. \ref{finalSU2LsMRamp}, the experimental curve for the real part
$R_0^0(s)$ of $T_0^0(s)$ is plotted up to about $\sqrt{s} = 1.2$ GeV.
Predictions for $m_{\sigma b}= 0.5,0.8$ and 1.0 GeV are also shown.
It is seen that the data up to roughly $\sqrt{s} = 0.8$ GeV can be fit
when $m_{\sigma b}$ lies in the 0.8 - 1.0 GeV range.  Thus, perhaps
surprisingly, the simple model does a reasonable job of accounting for
the low energy, but not just the linear threshold region, s-wave $\pi
\pi$ scattering data.  This circumstance enhances the plausibility of
the model based on the Lagrangian of Eq. (\ref{LsMLag}) together with
the unitarization scheme of Eq. (\ref{Tregularization}).

\begin{figure}[htbp]
\vspace{0.5cm}
\begin{center}
\mbox{\epsfig{file=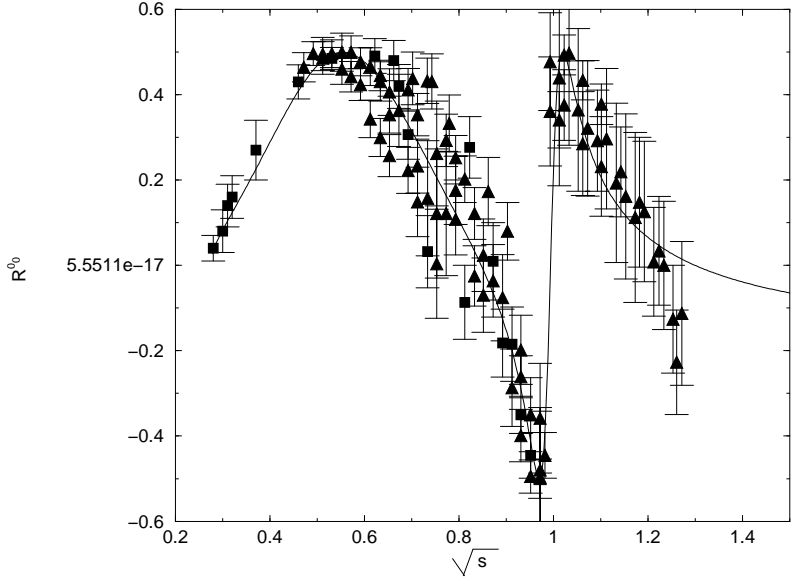,height=10cm,angle=270}}
\end{center}
\caption[]
{Comparison of the best fit for the Real part of the I=J=0 $\pi \pi$ scattering
amplitude in the non-renormalizable SU(3) Linear Sigma Model with
experiment.  Figure taken from \cite{BFMNS01}.  Best fit value of the bare $\sigma$ meson mass,
$m_{\sigma b}$ is 0.85 GeV.}
\label{finalSU3LsMNRpipiamp}
\end{figure}

There still may be some lingering doubt because the energy region
between about 0.8 and 1.2 GeV is not at all fit by the model.  However
this is due to the neglect of a second scalar resonance which is
expected to exist in low energy QCD.  As shown recently in
\cite{BFMNS01} if the Lagrangian of Eq. (\ref{LsMLag}) is ``upgraded''
to the three-flavor case \cite{Levy}-\cite{CH}
 (so that another scalar field $\sigma^\prime$
identifiable with the $f_0(980)$ is contained)
 the entire region up to about $\sqrt{s} = 1.2$ GeV can
be reasonably fit.  This is shown in Fig. \ref{finalSU3LsMNRpipiamp};
  in obtaining this fit
the value $m_{\sigma b} = 0.847$ GeV was selected.  Two other
parameters, $m_{\sigma^\prime b} = 1.300$ GeV and a $\sigma -
\sigma^\prime$ mixing angle were also determined in the fit.  Most
importantly for our present purpose, exactly the same calculational
scheme of simply feeding the tree approximation into the unitarization
formula of Eq. (\ref{Tregularization}) was employed.

In assessing the validity of this approximate model for low energy QCD
one should also consider the role of the vector mesons.  These are
known to be important in many low energy processes and give
the dominant contributions to the ``low energy constants'' \cite{leconstants}
 of the
chiral perturbation theory expansion.  Nevertheless it was found
\cite{HSS2}  that, while rho meson
 exchange does make a contribution to low energy s-wave pion pion
 scattering, its inclusion
does not qualitatively change the properties of the light $\sigma$
resonance which seems crucial to explain the I=J=0 partial wave.  More
specifically, the effect of the rho meson raises the $\sigma$ mass by
about 100 MeV and lowers its width somewhat. 

It also seems worthwhile to remark that this unitarized linear
 sigma model approach gives similar results to a more conventional non-linear
sigma model approach \cite{HSS1} wherein the scalar mesons are included 
explicitly. In order to get the effects of the vector and axial vector 
mesons in the linear sigma model framework one should also add them
explicitly in a chiral symmetric manner (e.g. \cite{H72}).

 \section{Sigma pole position}

Let us ask:  how non-perturbative is the linear sigma model when it is
employed as an approximation to low energy QCD?  The ordinary
criterion for the model to be deep in the non-perturbative region is
that the dimensionless coupling constant $b$ in Eq. (\ref{LsMLag}) be
very much greater than unity.  Using Eqs. (\ref{Fpi}) and (\ref{b})
this criterion reads

\begin{equation}
b = { \left( \frac{m_{\sigma b}}{2F_\pi} \right) }^2 = { \left
( \frac{m_{\sigma b}}{2 \sqrt{2} v} \right) }^2 \gg 1.
\end{equation}

Taking $m_{\sigma b} \approx 0.85$ GeV to fit experiment, as discussed
in the last section, gives a value $b=10.5$.  Thus it seems fair to
say that the theory lies outside the perturbative region \cite{DW89}.
Nevertheless, the K-matrix unitarization can lead to a result in
agreement with experiment.  In a non-perturbative regime one might
expect the physical parameters like the sigma mass and width to differ
from their ``bare'' or tree-level values.  To study this we look at
the complex sigma pole position in the partial wave amplitude in
Eqs. (\ref{Tregularization}) and (\ref{alphabeta}):  

\begin{equation}
T_0^0(s) = \frac { (m_{\sigma b}^2 - s) \alpha (s) + \beta(s) }
{ (m_{\sigma b}^2 - s)[1 - i \alpha(s)] - i \beta(s)}.
\label{Tunitary}
\end{equation}
This is regarded as a function of the complex variable $z$ which
agrees with $s + i \epsilon$ in the physical limit.  The pole position
$z_0$ is then given as the solution of:

\begin{equation}
(m_{\sigma b}^2 - z_0)[ 1 - i \alpha(z_0)] - i \beta(z_0) = 0.  
\label{poleeq}
\end{equation} 
Note that $\alpha(s)$ remains finite as $q^2=s-4m_\pi^2 \rightarrow
0$, so there are no poles due to the numerator of Eq. (\ref{Tunitary}).

For treating both the low energy QCD as well as the standard
electroweak situation it is convenient to introduce the scaled quantities :
\begin{eqnarray}
{\bar m} &=& \frac{ m_{\sigma b} }{F_\pi} = \frac{ m_{\sigma b} }{
\sqrt{2} v} \nonumber \\
{\bar z_0} &=& \frac{z_0}{F_\pi^2} = \frac {z_0}{2v^2}.
\end{eqnarray}

Then, specializing to the unbroken $SU(2) \times SU(2)$ case by 
setting
$m_\pi^2 = 0$ in Eq. (\ref{alphabeta}) we may write Eq. (\ref{poleeq})
explicitly as 
\begin{equation}
( {\bar m}^2 - {\bar z_0}) \{ 1 - i \frac{{\bar m}^2}{32 \pi}
\left[ -10 + \frac{4 {\bar m}^2}{\bar z_0} {\rm ln} ( 1 + \frac {\bar
z_0}{{\bar m}^2}) \right] \} - \frac{3i {\bar m}^4}{16 \pi}
= 0.  
\label{scaledvarpoleeq}
\end{equation}

It is easy to solve this equation numerically.  In Fig. 3 we show how
$\sqrt {{\rm Re} (z_0)}$ depends on the choice of $m_{\sigma b}$ for low
energy QCD (solid line).  For comparison the situation with $m_\pi =
0.137 $ GeV is also shown (dashed line).  We note that the behavior is
qualitatively similar.    

\begin{figure}[htbp]
\vspace{0.5cm}
\begin{center}
\mbox{\epsfig{file=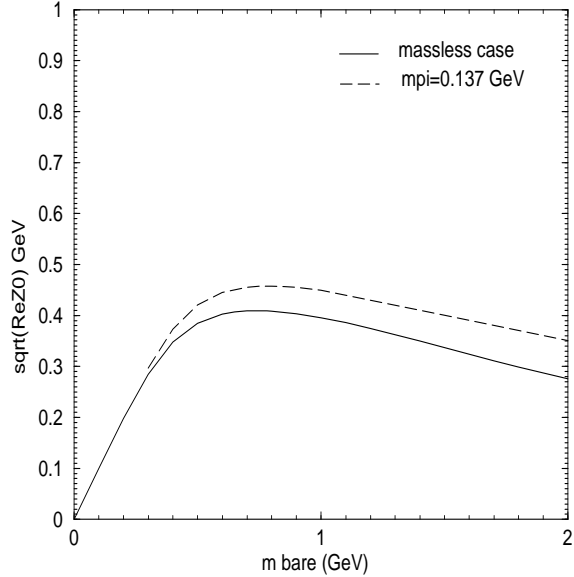, height=3in,width=3in, angle=0}}
\end{center}
\caption
{Plot of $\sqrt {{\rm Re} (z_0)}$ in GeV as a function of $m_{\sigma b}$
in GeV for the QCD application.    }
\label{LsMpoleSqrtRez0}
\end{figure}

In Fig. 4, we show how $\sqrt {-{\rm Im} (z_0)}$ depends on the choice
of $m_{\sigma b}$ for low energy QCD.  
 One sees that the real and imaginary pieces of $z_0$ are bounded.
Thus we can get an accurate analytic approximation to $z_0$ for large
$m_{\sigma b}$ by expanding the ``log'' in Eq. (\ref{scaledvarpoleeq}).  In
order to get the leading order approximation to $ \bar {z_0}$ it is
necessary to keep only three terms:  $ \ln (1 + x) \approx x -
\frac{1}{2} x^2 + \frac{1}{3}x^3$.  Then, in the $SU(2) \times SU(2)$
invariant case with $m_\pi = 0$ 
\begin{equation}
{\bar z_0} \approx \frac{352}{3} \frac{\pi^2}{{\bar m}^2} - 8 \pi i, 
\label{approxpoleeq}
\end{equation}
at large $ {\bar m} = \frac{m_{\sigma b}}{\sqrt{2} v}$ .

\begin{figure}
\vspace{0.5cm}
\begin{center}
\mbox{\epsfig{file=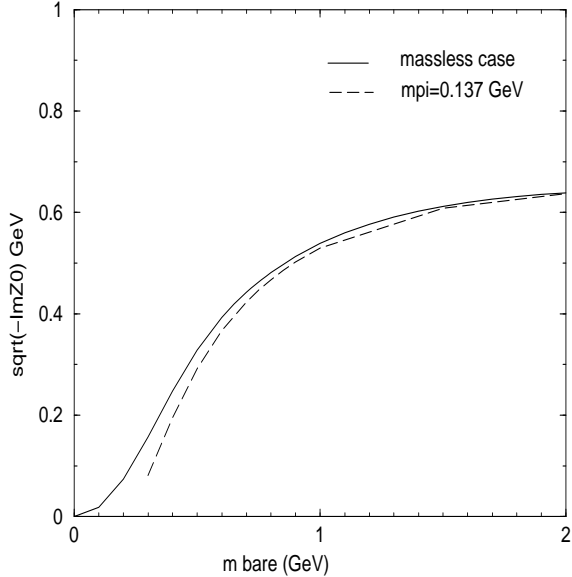, height=3in,width=3in, angle=0}}
\end{center}
\caption
{Plot of $\sqrt {-{\rm Im} (z_0)}$ in GeV as a function of $m_{\sigma b}$
in GeV for the QCD application}
\label{LsMpoleSqrtmImz0}
\end{figure}

The quantity $\sqrt {{\rm Re} (z_0)}$ can be interpreted as
a physical (renormalized) mass. Now from Fig. \ref{LsMpoleSqrtRez0}
 and Eq. (\ref{approxpoleeq}) we note that $\sqrt {{\rm Re} (z_0)}$ decreases
monotonically to zero for large $\bar m$.  There is a maximal value of
the physical mass, $\sqrt {{\rm Re} (z_0)}$ around $m_{\sigma b} =
0.74(0.79)$ GeV in the QCD case with $m_\pi = 0 (0.137)$ GeV.  This
corresponds to the scaled variable $ \bar m$ taking the value 5.65
(6.03).  Increasing $ \bar m$ beyond this point will decrease the
physical mass.  Thus there are two distinct $\bar m$'s for each value
of the physical mass.  Low energy QCD selects,
  in the sense of fitting to $\pi \pi $
scattering, $m_{\sigma b}$ around 0.85 GeV (with $m_\pi = 0.137 $
GeV) and the corresponding $\bar m$ around 6.5.  The physical mass at
that point is around 0.46 GeV.  The same physical mass would also
arise when $m_{\sigma b}$ is about 0.75 GeV.  However, as may be seen
from Fig. \ref{LsMpoleSqrtmImz0}, $\sqrt {-{\rm Im} (z_0)}$ is different
for the two cases.  At small $m_{\sigma b}$ we may plausibly identify
the physical width of the sigma as 
\begin{equation}
\Gamma_{\rm phy} = \frac{-{\rm Im} z_0}{\sqrt{{\rm Re} z_0}}.  
\label{physwidth}
\end{equation}

For large $m_{\sigma b}$ we may regard this as a convenient measure of
the width, even though the simple Breit-Wigner approximation as we
will discuss in the next section, 
is doubtful there.  From
Eq. (\ref{approxpoleeq}) we note that, at large $m_{\sigma b}$
\cite{pionmass}
 the physical sigma mass goes to
zero proportionally to $ \frac{v^2}{m_{\sigma b}}$ while the physical 
measure of
the width in Eq. (\ref{physwidth}) increases as $m_{\sigma b}$.
Clearly, this limiting behavior is different from a narrow
Breit-Wigner resonance.  For a descriptive understanding of the
non-perturbative situation it is probably better to look at the
unitarized amplitude, Eq. (\ref{Tregularization}) itself.  The real
part is illustrated in Fig. \ref{finalSU2LsMRamp} and will be further
discussed in the next section.  The mathematical significance of the
pole in this non-perturbative situation is that the amplitude is
approximately given by the sum of a complex constant, $b_\sigma$ and
the pole term:
\begin{equation}
T_0^0(s) \approx b_\sigma + \frac{a_\sigma}{s - z_0}.
\end{equation}
Here the residue $a_\sigma$ is actually complex.  The numerical values
of $a_\sigma$ and $b_\sigma$ are given in Table I of \cite{BFMNS01} for the SU(2) linear sigma
model case.  A check of the accuracy of this approximation for the
more complicated SU(3) linear sigma model case is illustrated in
Fig. 9 of the same reference.  

\section{Application to the $SU(2) \times U(1)$ electroweak model}

Whereas the Lagrangian Eq. (\ref{LsMLag}), characterized by the scale
$ v = \frac{0.131}{\sqrt{2}}$ GeV, is supposed to be an effective theory
for calculating the J=0 partial wave amplitudes of low energy QCD, the same
Lagrangian [written as Eq. (\ref{HiggsLag})] is generally considered
to be a portion of the minimal electroweak Lagrangian,
characterized by the scale $ v = 0.246 $ TeV.  Of course, many people
consider this portion to be the least well established aspect of the
electroweak theory.  Furthermore, there is a reasonable probability
that the standard model itself is part of a still larger theory.  Thus
it may be appropriate to regard the Higgs sector, Eq. (\ref{HiggsLag}),
unitarized by the K matrix approach, as a kind of effective
prescription.  An alternative "canonical" way to proceed, which has
been intensively investigated \cite{DH89,DoR90} would be to replace it by the
non-linear effective chiral Lagrangian, improved by systematically
including loops and higher derivative terms.  However this procedure
is expected to be practical only to an energy below the Higgs mass
(e.g. up to about 0.45 GeV in the analog QCD model, as shown in
Fig. \ref{finalSU3LsMNRpipiamp}).  Clearly it is desirable to
consider a model, like the present one, which has the
possibility of describing the scattering amplitude around the energy
of the Higgs boson even if it were to exist in a non-perturbative
scenario.   

The discussion of the $\pi \pi$ scattering amplitude $T_0^0$ in
sections 3 and 4 can also be used to treat the high energy scattering
of the longitudinal components of the W and Z bosons in the electroweak
theory by making use of the Goldstone boson equivalence
 theorem \cite{CLT74,LQT77,CG85}.
This theorem implies that the vector boson scattering amplitudes are
related to the scattering of the Nambu-Goldstone bosons (with $m_\pi =
0$  ) of the electroweak theory as
\begin{eqnarray}
{\rm amp}(W_L^+ W_L^- \rightarrow W_L^+ W_L^-) &=& {\rm amp}(\pi^+ \pi^-
\rightarrow \pi^+ \pi^-)  + {\cal{O}} ( \frac{m_W}{E_W} )  \nonumber \\
{\rm amp}(W_L^+ W_L^- \rightarrow Z_L Z_L) &=& {\rm amp}(\pi^+ \pi^-
\rightarrow \pi^0 \pi^0)  + {\cal{O}} ( \frac{m_W}{E_W} )   \quad {\rm
etc.}
\label{GBEquiv}
\end{eqnarray}

\begin{figure}
\vspace{0.5in}
\begin{center}
\epsfig{file=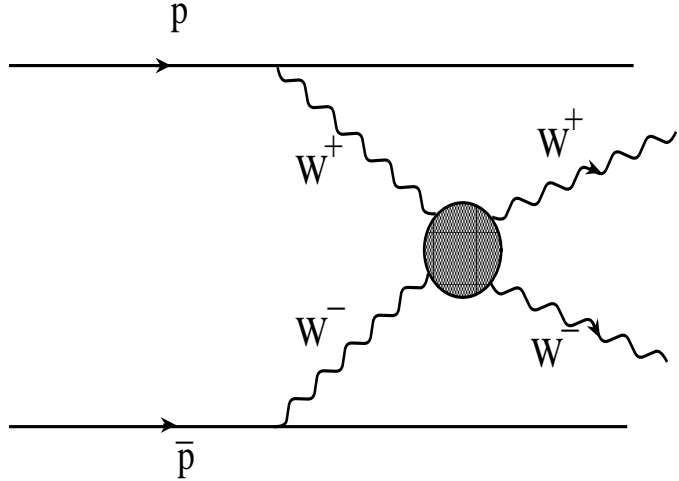, height=2.5in, width=3.5in, angle=0}
\end{center}
\caption
{Schematic illustration of $p {\bar p}\rightarrow W^+ W^- X$}
\label{ppbar}
\end{figure}

In the amplitudes on the right hand sides, the value
 $ v = \frac{F_\pi}{\sqrt{2}} = 0.246$ TeV should of course be used.
  We will specialize to
the $T_0^0(s)$ partial wave as in Eq. (\ref{Tregularization}) and will
set $m_\pi = 0$ in Eq. (\ref{alphabeta}).
This partial wave amplitude will contribute important pieces
to the reactions in Eq. (\ref{GBEquiv}), especially in the vicinity
 of the Higgs pole.  When folded together with the appropriate
 strong interaction pieces
the scattering is in principle \cite{DGMP97} measurable
 from $ p \bar p$ processes
like the one schematically illustrated in Fig. \ref{ppbar}.  
Of course, competing contributions must also be disentangled.

Now, we want to compare the unitarized longitudinal electroweak vector
boson scattering with the analog $ \pi \pi $ scattering.   Figs. 3 and
4 show that, in the latter case the effect of $m_\pi$ non-zero is
qualitatively small.  Since $\frac{m_W}{v_{\rm weak}} <
\frac{m_\pi}{v_{\rm QCD}}$ , the effect of non-zero $m_W$ is expected to be
even less significant.
 Then the Higgs pole positions can be gotten from Figs. 3
and 4 using the scaled quantities defined in Eq. (\ref{scaledvarpoleeq}) or
the asymptotic formula Eq. (\ref{approxpoleeq}).  However for
convenience we display the Higgs pole position in Fig. 6.

\begin{figure}
\vspace{0.5cm}
\begin{center}
\mbox{\epsfig{file=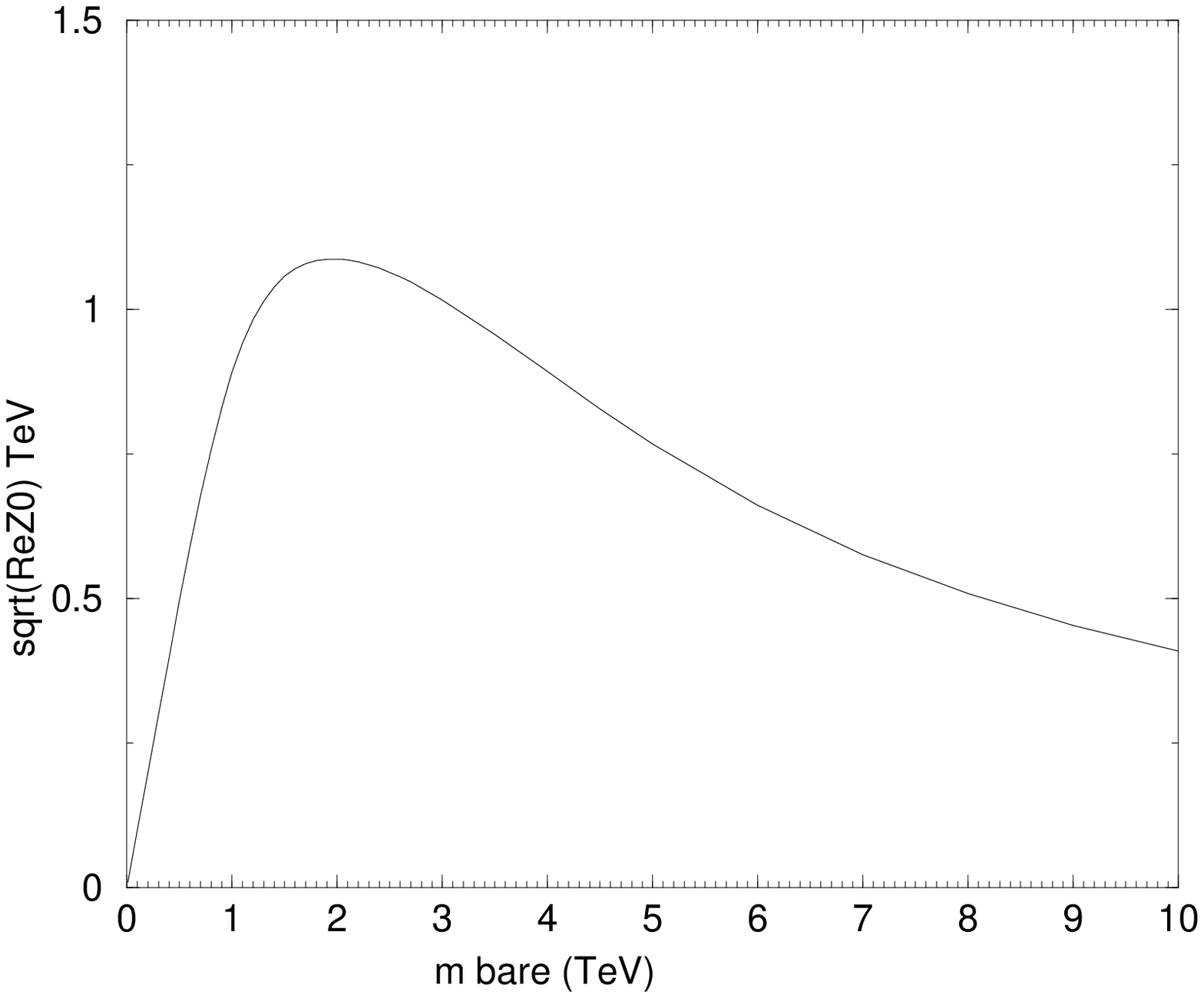, height=3in,width=3in, angle=0}}
\mbox{\epsfig{file=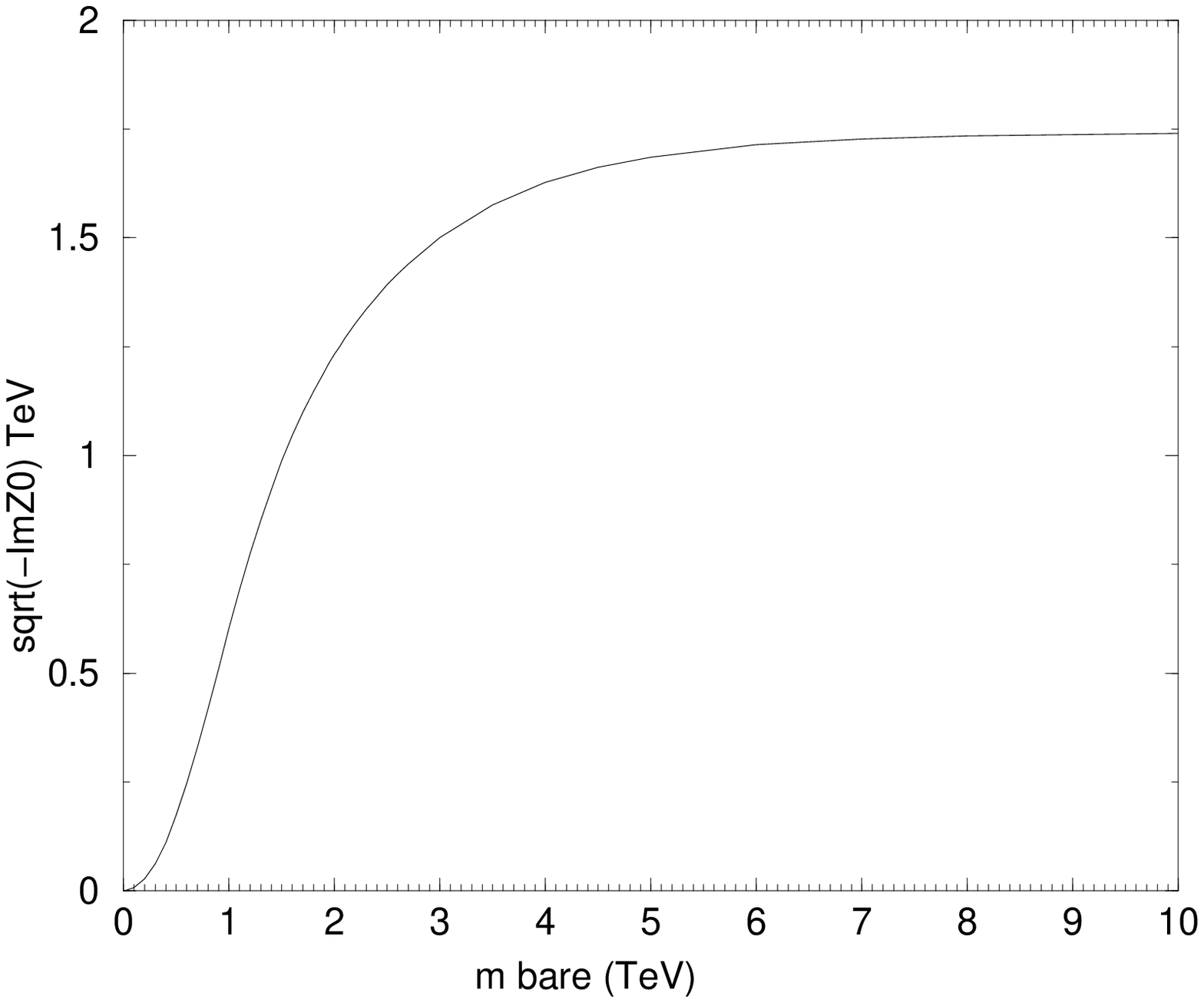, height=3in,width=3in, angle=0}}
\end{center}
\caption
{Square root of modulus of Real and Imaginary parts of pole in
unitarized amplitude as a function of bare Higgs mass. $\sqrt{ReZ_{pole}}$may
be identified as ``physical'' Higgs mass.  Physical width may be
identified from Eq. (\ref{physwidth}).}
\label{PhysicalHiggsMass}
\end{figure}

For orientation we
 recall that the ``QCD sigma'' has a bare mass
$m_{\sigma b}$ of about 850 MeV which gives a dimensionless mass $\bar
m = 6.49$ \cite{zerompi}.  This value of $\bar m$ corresponds
 to a bare Higgs mass value of $m_{\sigma
b} = 2.26$ TeV.  At that value, the measure of the physical Higgs
mass, $\sqrt{ {\rm Re}(z_0)}$ would be about 1.1 TeV and $\sqrt{ -{\rm
Im}(z_0)}$ would be about 1.3 TeV.  Evidently the QCD sigma is much
further in the non-perturbative region than the corresponding range of
values ($\leq$ several hundred GeV) which are now usually considered
 for the Higgs mass. 
 Of course there is no reason for the
scalar Higgs parameter $ \bar m_0$ to agree with the QCD value.  The
significance of the present observation is that it increases one's
confidence in the applicability of the K-matrix unitarization model
for Eq. (\ref{HiggsLag}) to bare Higgs masses in at least the 2 TeV
region.  In other words, the same model with the same scaled
parameters agrees with experiment there in a different context.  Since the present
status of the Higgs sector of the standard electroweak theory may
change if future experiments reveal evidence for new physics just
beyond the standard model, it seems worthwhile to have some confidence
in an approach to a possibly non-perturbative Higgs sector.  In
particular as the QCD analog two-sigma model discussed in section II
(see Fig. 2) shows, the K-matrix unitarization method can be expected
to work in the case of more than one Higgs particle even in
non-perturbative regions of parameter space. For example, the
introduction of the $\sigma'$ resonance which gives Fig. 2 would
scale to a 2 Higgs model describing physics up to about 3.3 TeV
in some electroweak theory.  

It may be interesting to give a brief survey of the characteristics of
the Higgs particle, as ``seen'' in the s-wave vector boson scattering
predicted by the unitarized Higgs model.  Figure
\ref{PhysicalHiggsMass} indicates that the physical Higgs mass,
$\sqrt{ {\rm Re}(z_0)}$, approximately agrees with the bare Higgs mass
$m_{\sigma b}$ until about $m_{\sigma b} = 0.5 $TeV.  Thereafter the
physical Higgs mass grows less quickly \cite{massdecrease}
 until it
peaks at about 1.1 TeV with $m_{\sigma b}$ about 2.0 TeV.  Afterwards,
$\sqrt{ {\rm Re}(z_0)}$ decreases gradually and vanishes as 
$\frac{v^2}{m_{\sigma b}}$ according to Eq. (\ref{approxpoleeq}).
  On the other hand, 
$\sqrt{- {\rm Im}(z_0)}$ monotonically increases to the saturation value
$4v \sqrt{\pi} = 1.74$ TeV.  This general pattern of pole position
vs. bare Higgs mass has been observed in a variety of non-perturbative
approaches \cite{CDG84,RS89,DR,W91}.  

At large values of $m_{\sigma b}$, the intuitive meaning of the pole
position is not immediately clear.  It may be more physical to
consider the question of the deviation of the predicted Higgs
resonance shape from that of a pure Breit Wigner resonance.  In Fig. 7
are plotted $ \left| T_0^0 (s) \right|$ as a function of $\sqrt {s}$
for various values of bare Higgs mass in comparison with the pure
Breit Wigner shape having the same bare mass and bare width
($\Gamma_{\rm bare} = \frac{3m_{\sigma b}^3}{32 \pi v^2}$).  In all
these plots, the Breit Wigner curve is higher before the peak and
lower after the peak.  Note that the actual {\it bare} amplitude in
Eqs. (\ref{pipiampl}) and (\ref{alphabeta}) includes the effect of
crossed channel Higgs exchanges and a contact term in addition to the
s-channel pole.  For a bare Higgs mass of 350 GeV there is still not
much deviation from the pure Breit Wigner shape.  However for
$m_{\sigma b} = 1$ TeV the deviation is rather marked.  In this case
the unitarized model appears realistic at lower energies before the
peak, which is not surprising since it obeys the low energy theorem
 so is forced to vanish as $s\rightarrow 0$.
  On the other
hand, the simple Breit Wigner looks unrealistically high at energies
below the peak due to its large width as discussed after
 Eq. (\ref{conventionalreg}). Above
the peak, however, the unitarized amplitude appears to rise and level
off.  This trend is clarified by the
plot for the $m_{\sigma b} = 3$ TeV case;  there the unitarized model
has a similar shape to the 1 TeV case below the peak
 but $ \left| T_0^0 (s) \right|$
simply saturates to unity for higher $\sqrt{s}$.  
It is also amusing to observe that the peak of
$|T^0_0(s)|$ always occurs at the bare Higgs mass, $s=m_{\sigma b}^2$. This
may be seen by noting that, since $T^0_0(s)$ may be expressed in
terms of the phase shift as ${\rm exp}[i\delta^0_0(s)]{\rm sin}\delta^0_0(s)$,
the peak will occur where $|\delta^0_0(s)|=\pi/2$.
In other words, the peak will occur where $T^0_0(s)$ is pure 
imaginary. This is immediately
 seen from Eq (\ref{Tunitary}) to be the case when $s=m_{\sigma b}^2$.

\begin{figure}
\vspace{0.5cm}
\begin{center}
\mbox{\epsfig{file=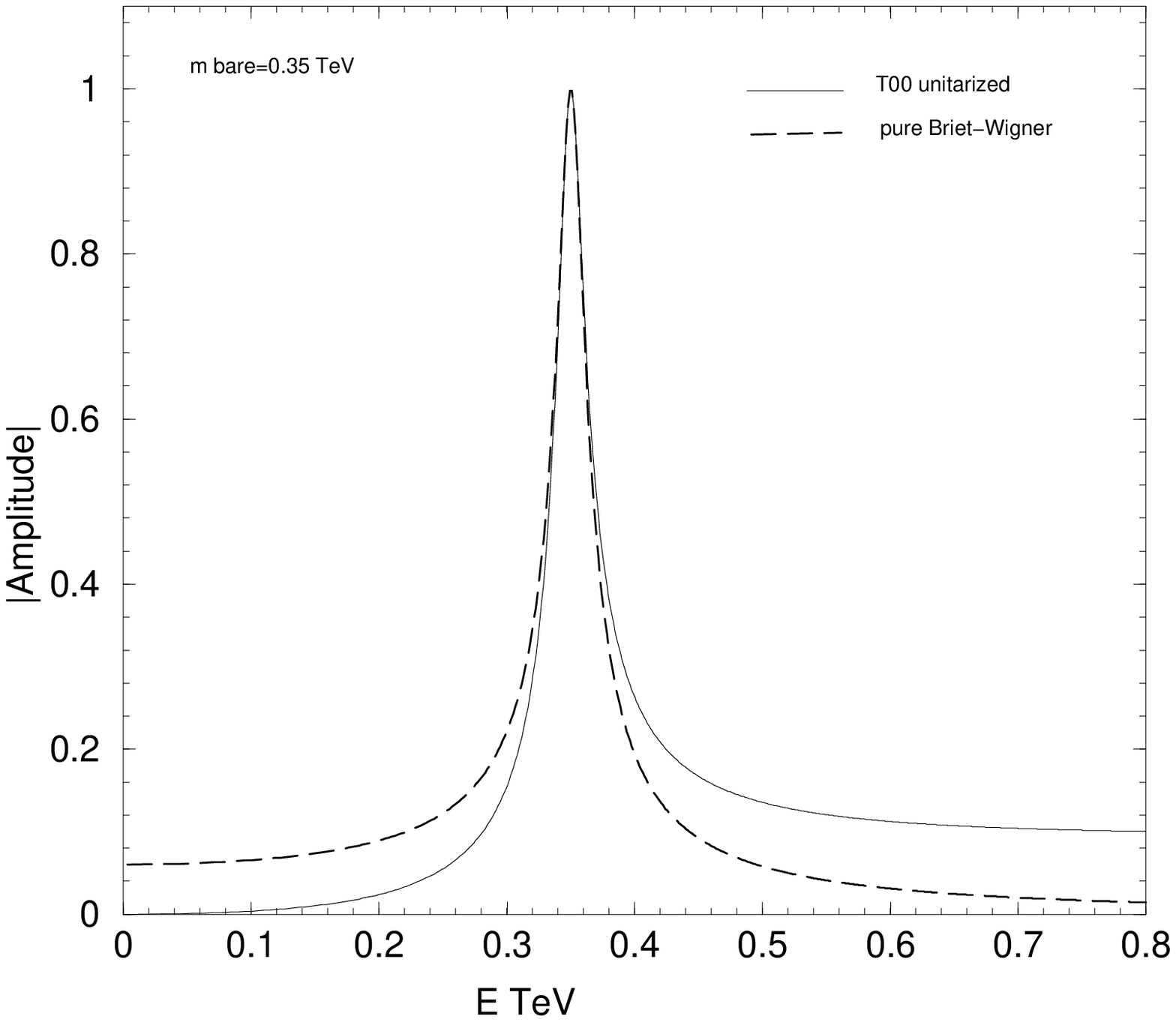, height=3in,width=3in, angle=0}}
\mbox{\epsfig{file=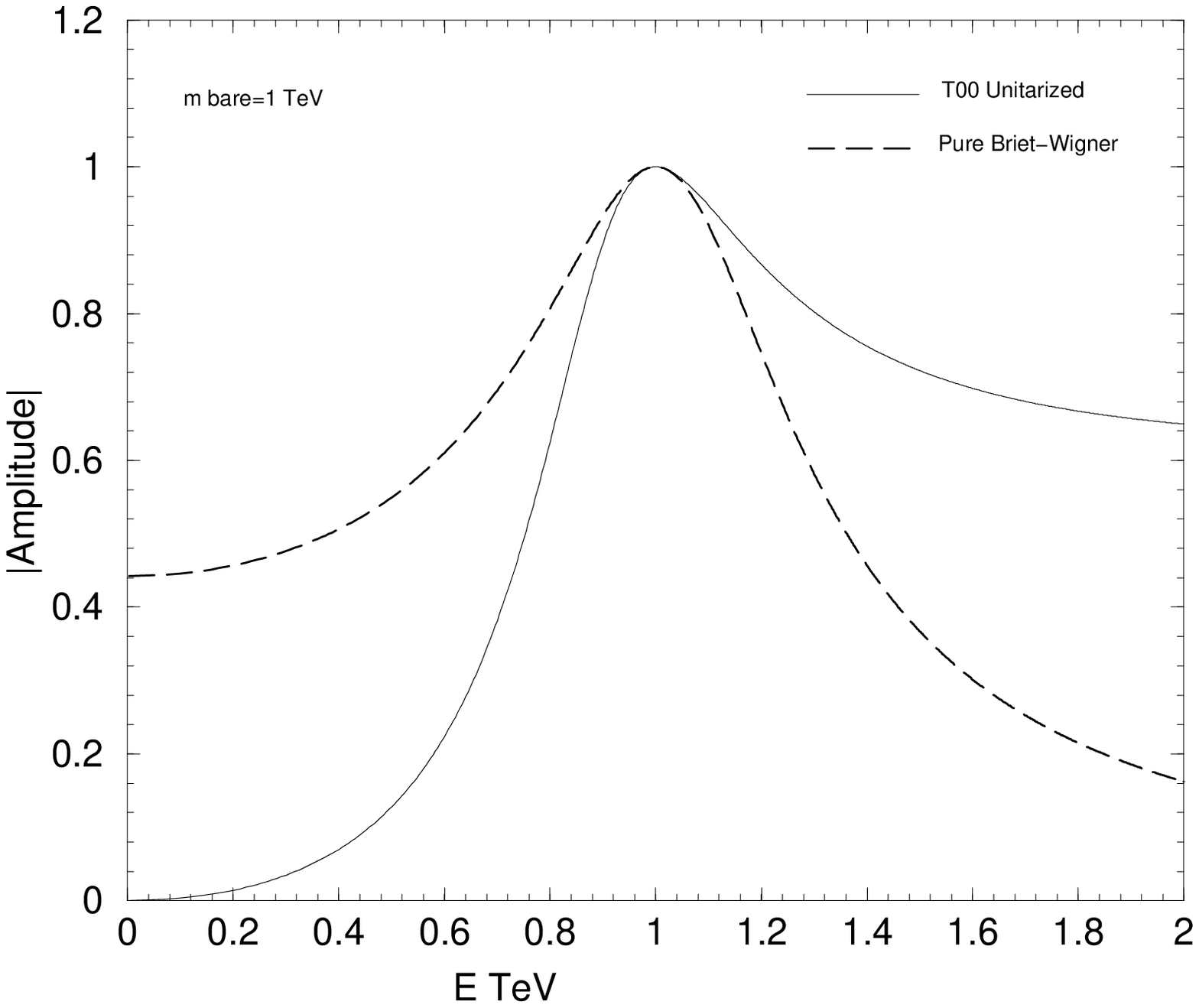, height=3in,width=3in, angle=0}}
\mbox{\epsfig{file=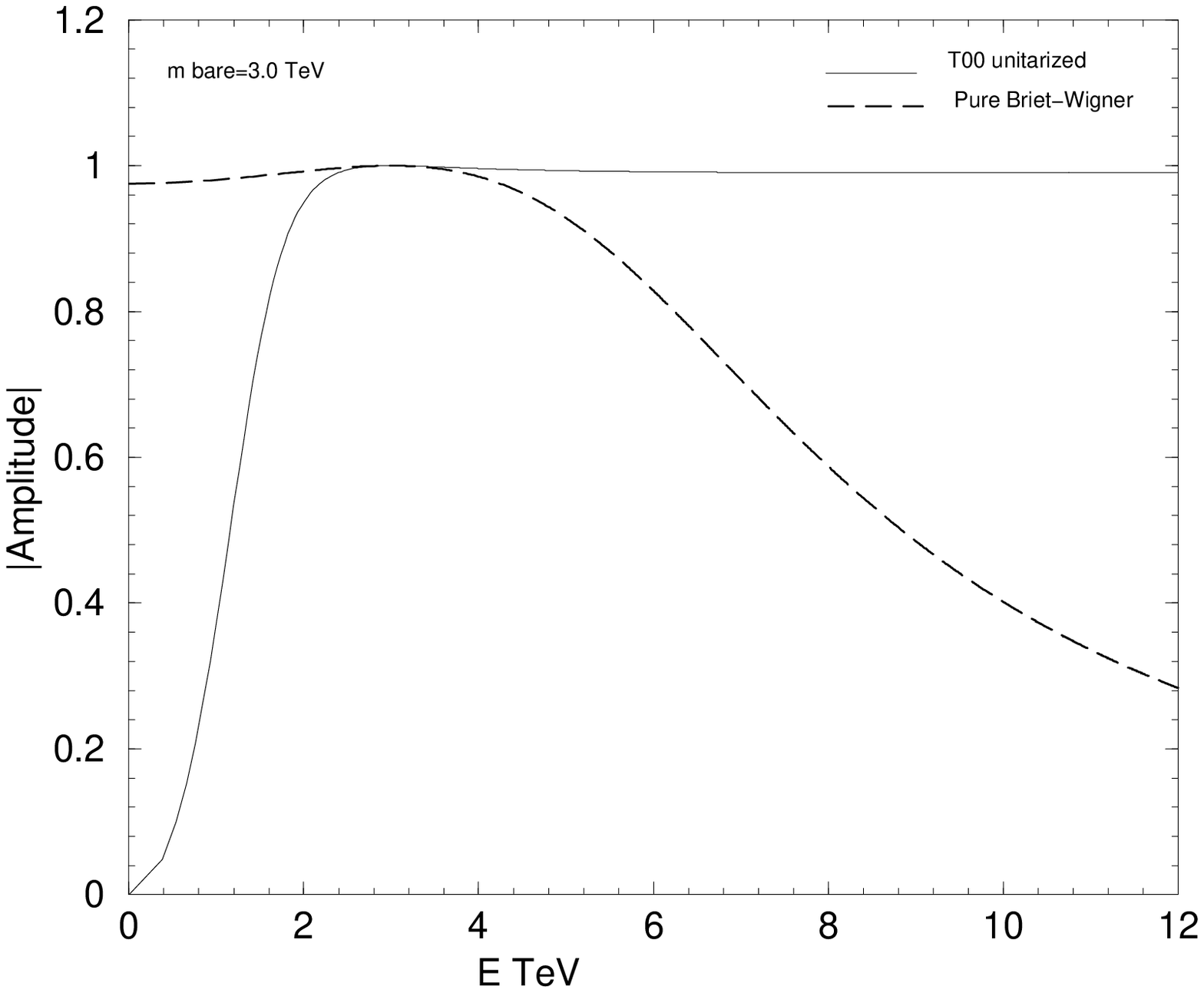, height=3in,width=3in, angle=0}}
\end{center}
\caption
{Comparison of absolute value of K-matrix unitarized (solid) and simple Breit
Wigner (dashed) I=J=0 amplitude for  $m_{\sigma b} = 0.35, 1, 3$ TeV.    }
\label{absHiggs_BW_350GeV}
\end{figure}

In order to better understand the large $s$ behavior of the unitarized
amplitude it seems helpful to examine the real and imaginary parts
${\rm Re} [T_0^0(s)]$ and ${\rm Im} [T_0^0(s)]$, for various values of
$m_{\sigma b}$.  Fig. 8 shows these for the same values of the bare
Higgs masses as in Fig. 7.  Notice that the real part ${\rm Re}
[T_0^0]$ always vanishes at $s = {m_{\sigma b}}^2$ while ${\rm Im}
[T_0^0]$ is always unity at that point.  [These features are evident on
inspection of Eq. (\ref{pipiampl}) together with
 Eq. (\ref{Tregularization}).]  While this aspect of a simple Breit-Wigner
resonance is preserved it is seen that the symmetry about the bare
mass point gets to be strongly distorted as the bare mass increases.

\begin{figure}[htbp]
\vspace{0.5cm}
\begin{center}
\mbox{\epsfig{file=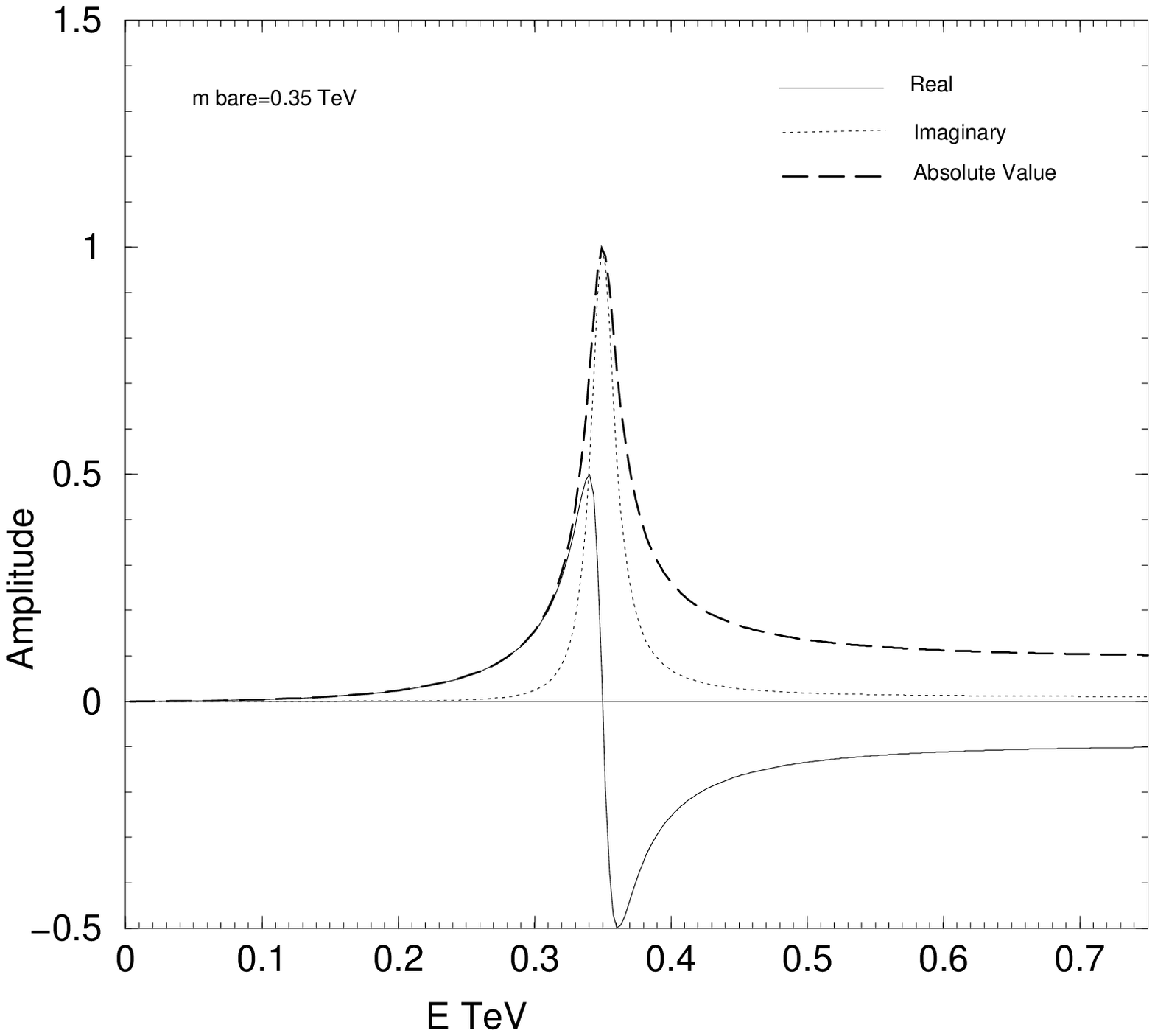, height=3in,width=3in, angle=0}}
\mbox{\epsfig{file=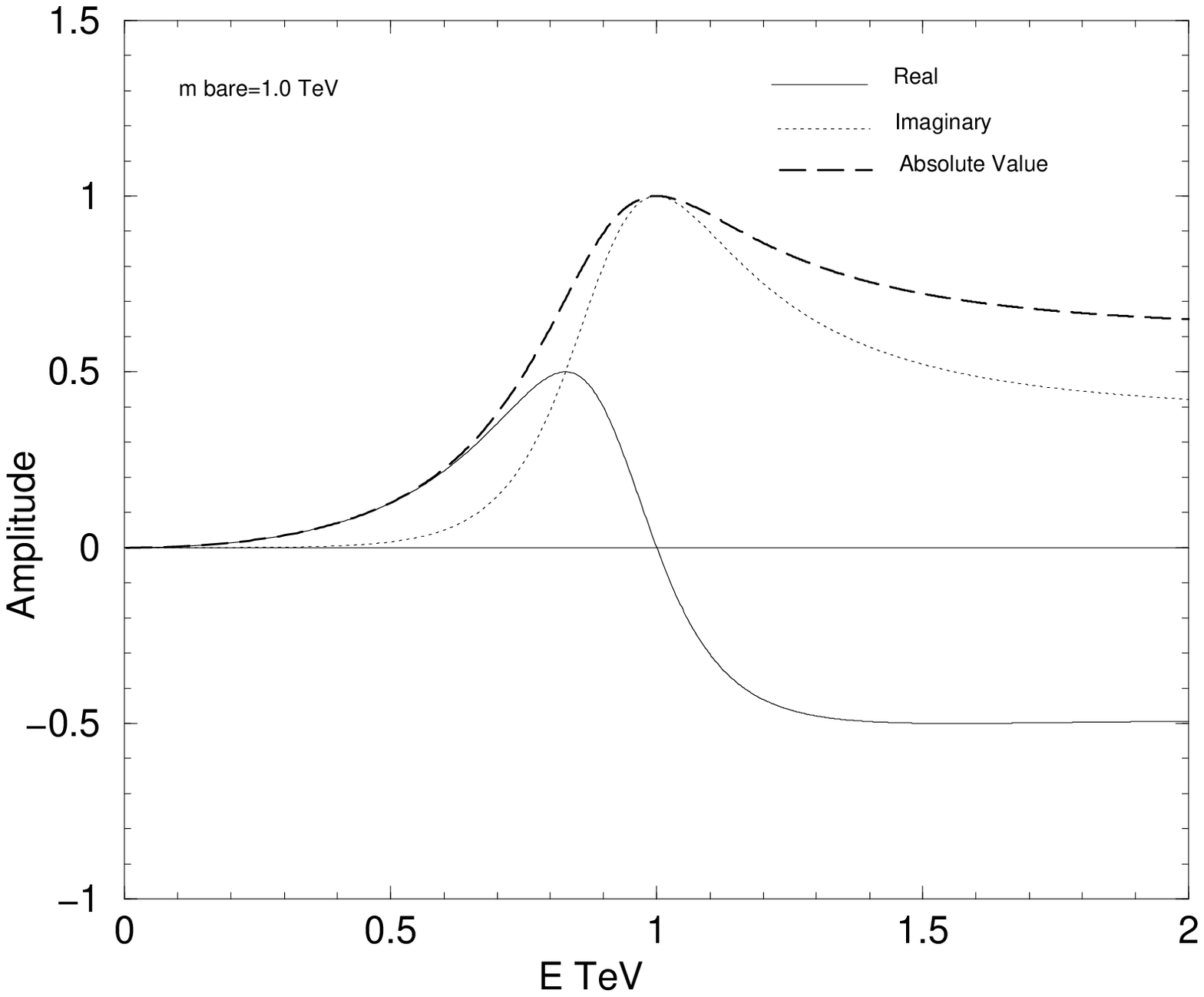, height=3in,width=3in, angle=0}}
\mbox{\epsfig{file=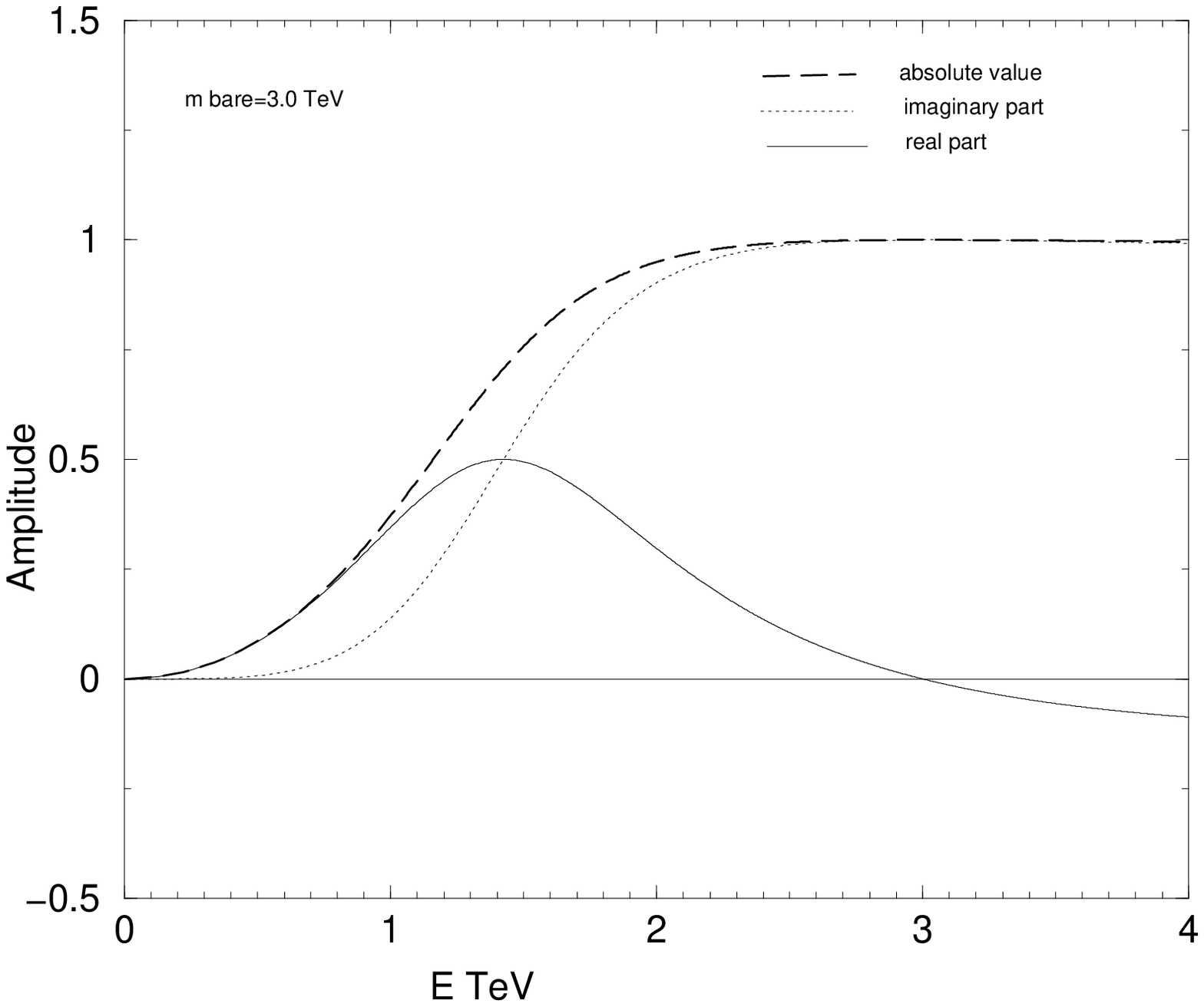, height=3in,width=3in, angle=0}}
\end{center}
\caption
{Real (solid), Imaginary (dotted) and Absolute value (dashed) of the 
K-matrix
 unitarized I=J=0 amplitude for  $m_{\sigma b} = 0.35, 1, 3$ TeV.    }
\label{ReImHiggs}
\end{figure}

One notices that both the real and imaginary parts flatten out at
large $s$ for all three choices of $m_{\sigma b}$.  This effect may be
described analytically by making a large $s$ expansion (with $m_\pi =
0$) for fixed bare mass $m_{\sigma b}$ of Eqs. (\ref{pipiampl}) and
(\ref{Tregularization}).  ${[T_0^0(s)]}_{\rm tree}$ becomes the constant
${{-5 {m_{\sigma b}}^2}} \over { 32 \pi v^2}$ which implies the flat
large $s$ behaviors

\begin{eqnarray}
{\rm Re}T_0^0 &\sim& \frac { \frac{-5 {m_{\sigma b}}^2} { 32 \pi
v^2} } { 1 + {( \frac{ 5{m_{\sigma b}}^2 }{32 \pi v^2})}^2 } \nonumber
\\
{\rm Im}T_0^0 &\sim& \frac{1}{ 1 + {( \frac{32\pi v^2}{5 m_{\sigma b}^2})}^2 }.
\end{eqnarray}

By construction, the amplitude is unitary for all $s$. 
It is clear that the amplitudes for these values of $m_{\sigma b}$
 vary significantly with energy only
 up to slightly above 
$m_{\sigma b}$.  The
physics beyond this point might be filled out if a heavier Higgs meson
exists and one would get a picture something like
Fig. \ref{finalSU3LsMNRpipiamp}.  Adding the effects of other
interactions involving the Higgs meson to ${[T_0^0(s)]}_{\rm tree}$ in
Eq. (\ref{Tregularization}) might also be expected to fill out the flat
energy region.  

From the point of view that the K-matrix unitarized model is
interpreted as an effective theory (which is unitary for any
$m_{\sigma b}$) it is especially interesting to consider the case
where $m_{\sigma b}$ gets very large.  As a step in this direction,
Fig. 9 shows a plot of the amplitude for $m_{\sigma b} = 10$ TeV.  It
is seen to be similar to the amplitude for $m_{\sigma b} = 3$ TeV,
except that the real part (which does go through zero at $\sqrt{s} =
10$ TeV) saturates to a value which is almost indistinguishable from
the horizontal axis.  In fact this is tending toward a ``universal''
curve - any larger value of $m_{\sigma b}$ will give a very similar
shape.  This may be seen by expanding the amplitude for large
$m_{\sigma b}$ while keeping $s \ll m_{\sigma b}$.  Then
${[T_0^0(s)]}_{\rm tree} \approx \frac{s}{16 \pi v^2} + {\cal
O}(\frac{s^2}{v^2 m_{\sigma b}^2})$ so, with
K matrix regularization, we get for large $m_{\sigma b}$

\begin{eqnarray}
{\rm Re}T_0^0(s) &\sim & \frac  {\frac{s}{16 \pi v^2}}{ 1 +
{(\frac{s}{16 \pi v^2})}^2} \nonumber \\ 
{\rm Im}T_0^0(s) &\sim & \frac {1} { 1 + {(\frac{16 \pi v^2 }{s})}^2}.
\label{unlsm}
\end{eqnarray}
 
 To show the trend towards saturation we have plotted in
 Fig. \ref{absHiggsKmatrix_various_mbare_4TeV} $|T^0_0(s)|$ for the bare Higgs
masses 1, 3, 10, $\infty$  TeV. What is happening should not be surprising.
As $m_{\sigma b} \rightarrow \infty$ the tree amplitude is going to nothing but
the "current algebra" one, $s/(16\pi v^2)$. This may be obtained directly
from the non-linear SU(2) sigma model, which is expected since the non-linear model
was originally \cite{GL} motivated by taking the
sigma bare mass, $m_{\sigma b}$ to be very large and eliminating
 the sigma field by its 
equation of motion. Thus Eqs. (\ref{unlsm}) just represents the K matrix
unitarization of the "current algebra" amplitude. Since the result is unitary
it is not a priori ridiculous to contemplate the possibility that
Eqs. (\ref{unlsm}) could be a reasonable representation of the physics
when $m_{\sigma b}$ describes a far beyond the standard model sector
of a more fundamental theory. However, all the
structure in the scattering amplitude is confined to the much lower 
energy range centered 
around $4v\sqrt{\pi} \approx$ 1.74 TeV. Of course it may be more likely
that the model would represent a unitarization of the theory
with bare Higgs mass in the 0.1-3 TeV range \cite{Pa00}.
 In any event, the simple
$m_{\sigma b} \rightarrow \infty$ limit nicely explains
 the evolution of the scattering 
amplitudes for large bare Higgs mass. We should also remark
that the non-linear sigma model can be more generally motivated
directly; there is no need to integrate out the sigma from
a linear model. This would lead to the alternative "canonical"
approach mentioned at the beginning of this section.

\begin{figure}[htbp]
\vspace{0.5cm}
\begin{center}
\mbox{\epsfig{file=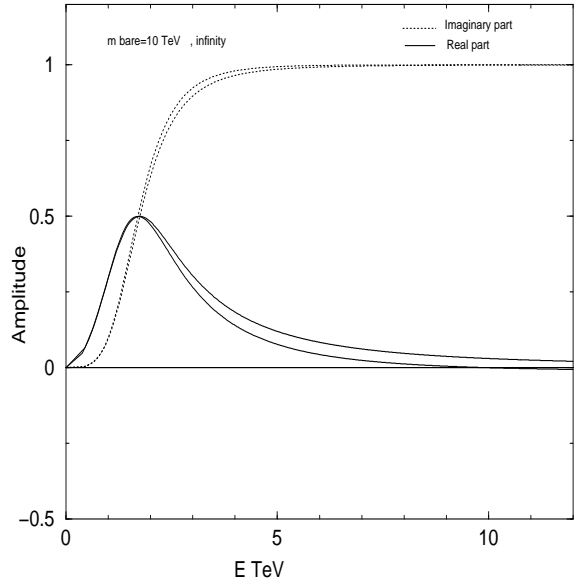, height=3in,width=3in, angle=0}}
\end{center}
\caption
{Real (solid) and Imaginary (dotted) parts of the K-matrix unitarized 
I=J=0 amplitude for  $m_{\sigma b} = 10$ TeV and $m_{\sigma b}=\infty$    
}
\label{ReImHiggs_10TeV}
\end{figure}

\begin{figure}[htbp]
\vspace{0.5cm}
\begin{center}
\mbox{\epsfig{file=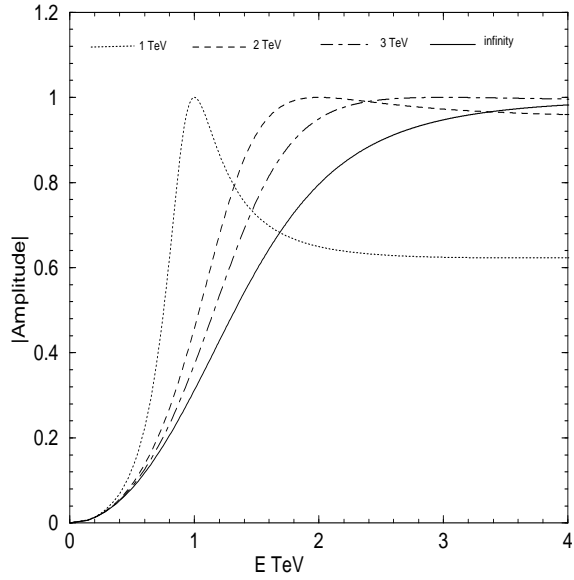,
height=3in,width=3in, angle=0}}
\end{center}
\caption
{Comparison of absolute value of unitarized amplitude $\left| T_0^0 \right|$ for
$m_{\sigma b} = 1(dotted),2(dashed),3(dash-dotted),\infty(solid)$ TeV. }
\label{absHiggsKmatrix_various_mbare_4TeV}
\end{figure}

\section{Additional discussion of the electroweak model}
\begin{figure}[htbp]
\vspace{0.5cm}
\begin{center}
\mbox{\epsfig{file=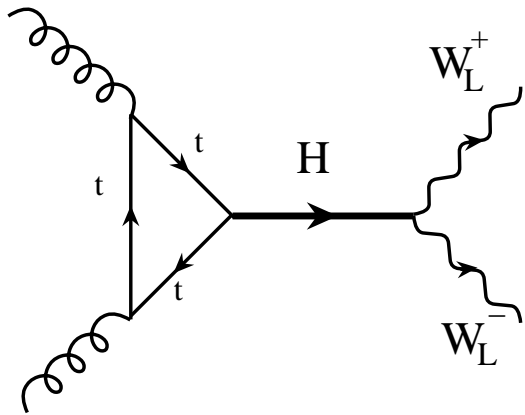,
height=7cm, angle=0}}
\end{center}
\caption
{Gluon Fusion Reaction }
\label{glufu}
\end{figure}

We just discussed the $W_L - W_L$ scattering making use of the simple
K-matrix unitarization and the Equivalence Theorem.  This has an
application  to the ``W-fusion'' reaction shown in Fig. \ref{ppbar}.
Now we point out that a similar method can be used to
 take into account strong final state interactions in the
``gluon-fusion'' reaction \cite{GGMN78} schematically illustrated in
Fig. \ref{glufu}. This is an interesting reaction since 
  it is predicted
\cite{GHPTW87,D88,GvdB89,KD91} that  
gluon fusion will be an important source of Higgs production.
 The t-quark shown running around the triangle makes
the largest contribution because the quarks, of course, couple to the
Higgs boson proportionally to their masses.   According to the
Equivalence Theorem, at high energies this Feynman diagram will contain a factor
${{g_{\sigma \pi \pi}} \over { ( m_{\sigma b}^2 - s )}}$ where the
$\pi$'s correspond to the $W_L$'s and $g_{\sigma \pi \pi} =
\frac{m_{\sigma b}^2}{v}$  appears in the trilinear
interaction term $\frac{g_{\sigma \pi \pi}}{2}{\sigma} {\mbox{\boldmath ${\pi}$}}\cdot 
{\mbox{\boldmath ${\pi}$}}$ obtained from Eqs. (\ref{LsMLag}) and (\ref{b}).  The need for unitarization is signaled,
as in the WW scattering case by the fact that this diagram has a pole
at $ s = m_{\sigma b}^2$.  Now it is necessary to regularize a three
point rather than a four point amplitude; this is discussed in
\cite{AS94,Chung}.  The WW scattering amplitude ${[T_0^0]}_{\rm tree}$
 in the previous section was unitarized by replacing
\begin{equation}
{[T_0^0]}_{\rm tree} \longrightarrow \frac{ {[T_0^0]}_{\rm tree}   }
{ 1 - i{[T_0^0]}_{\rm tree} } = {[T_0^0]}_{\rm tree} \left(   1 +
i{[T_0^0]}_{\rm tree} - {[T_0^0]}_{\rm tree}^2 + \ldots  \right).
\label{bubble}
\end{equation}

This has the structure of a bubble sum with the intermediate "Goldstone
bosons" on their mass shells.  For the gluon fusion reaction the final
"Goldstone bosons" similarly rescatter and we should replace
\begin{eqnarray}
\frac{  g_{\sigma \pi \pi} }{m_{\sigma b}^2 - s} &\longrightarrow&
\frac{  g_{\sigma \pi \pi} }{m_{\sigma b}^2 - s} \frac{ 1 }
{ 1 - i{[T_0^0]}_{\rm tree} } \nonumber \\
&=& \frac{  g_{\sigma \pi \pi} }{m_{\sigma b}^2 - s} \cos \delta_0^0
{\rm e}^{i \delta_0^0}.
\label{FSI}
\end{eqnarray}
In the second step we used the relation for the phase
shift $\delta_0^0$ in the K-matrix unitarization scheme:  $\tan
\delta_0^0 = {[T_0^0]}_{\rm tree}$, where the unitarized I=J=0 S-matrix
element is given by $S_0^0 = \exp{2i\delta_0^0}$. 

\begin{figure}[htbp]
\vspace{0.5cm}
\begin{center}
\mbox{\epsfig{file=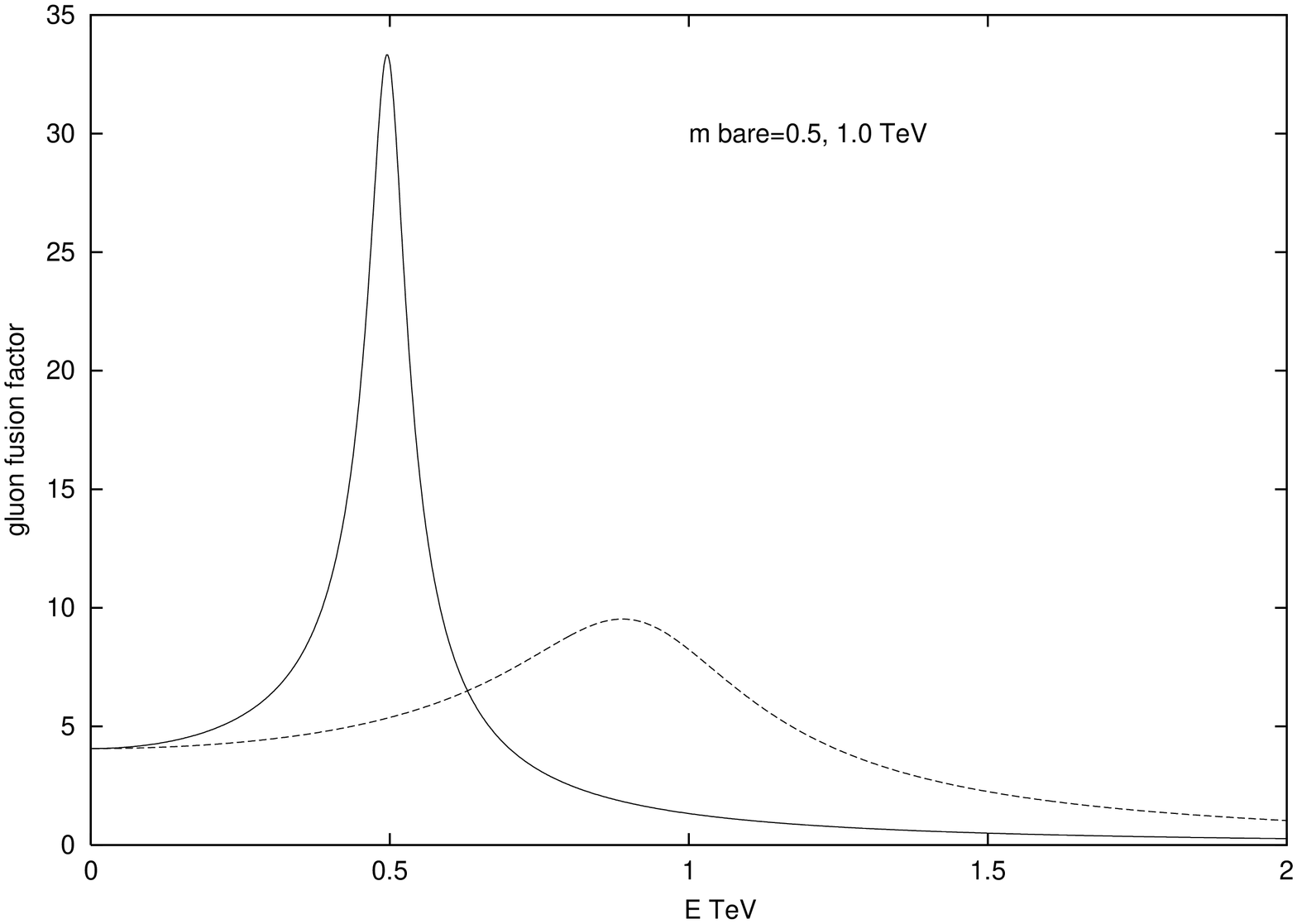,height=3in,width=3in,angle=270}}
\mbox{\epsfig{file=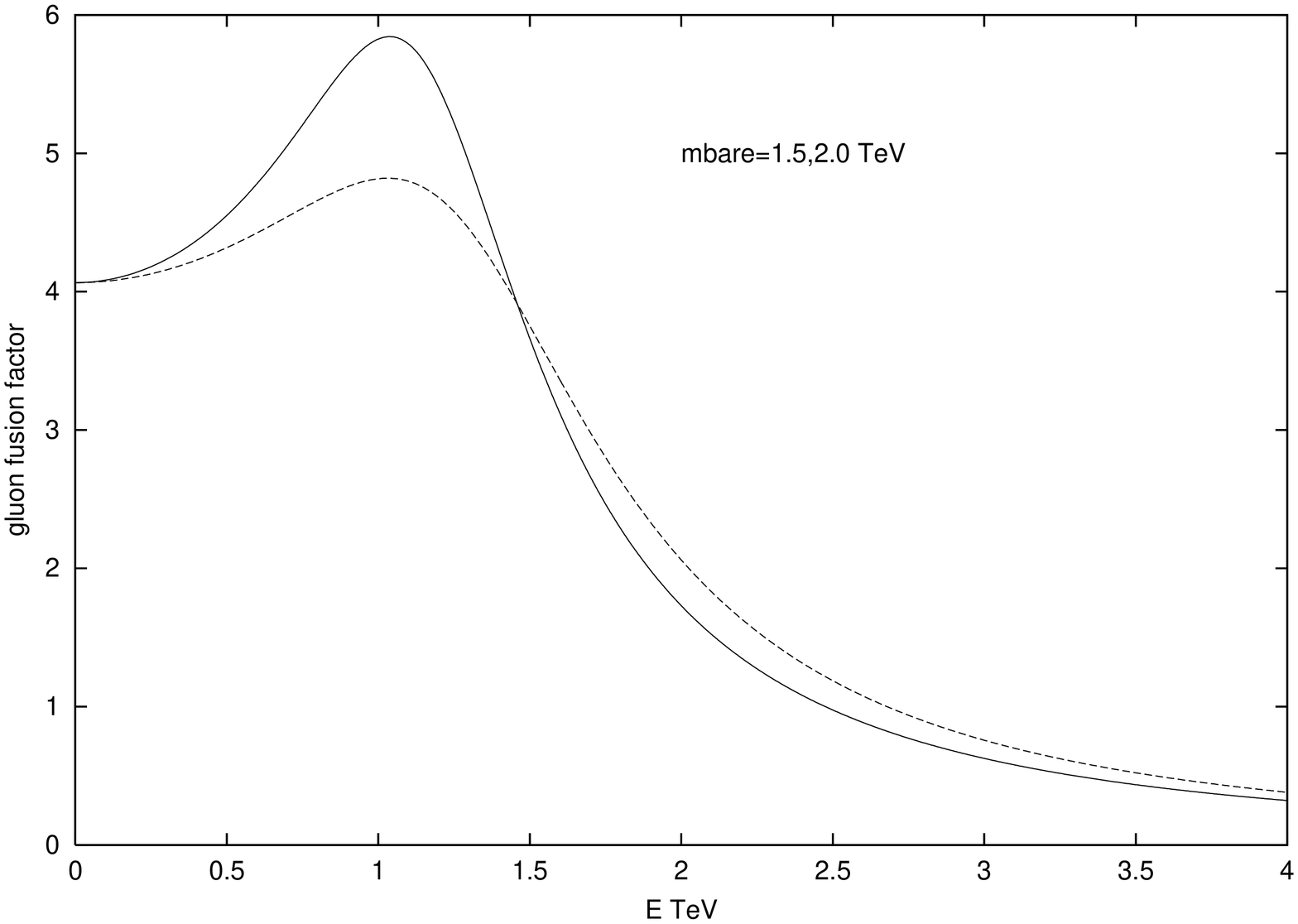,height=3in,width=3in, angle=270}} 
\mbox{\epsfig{file=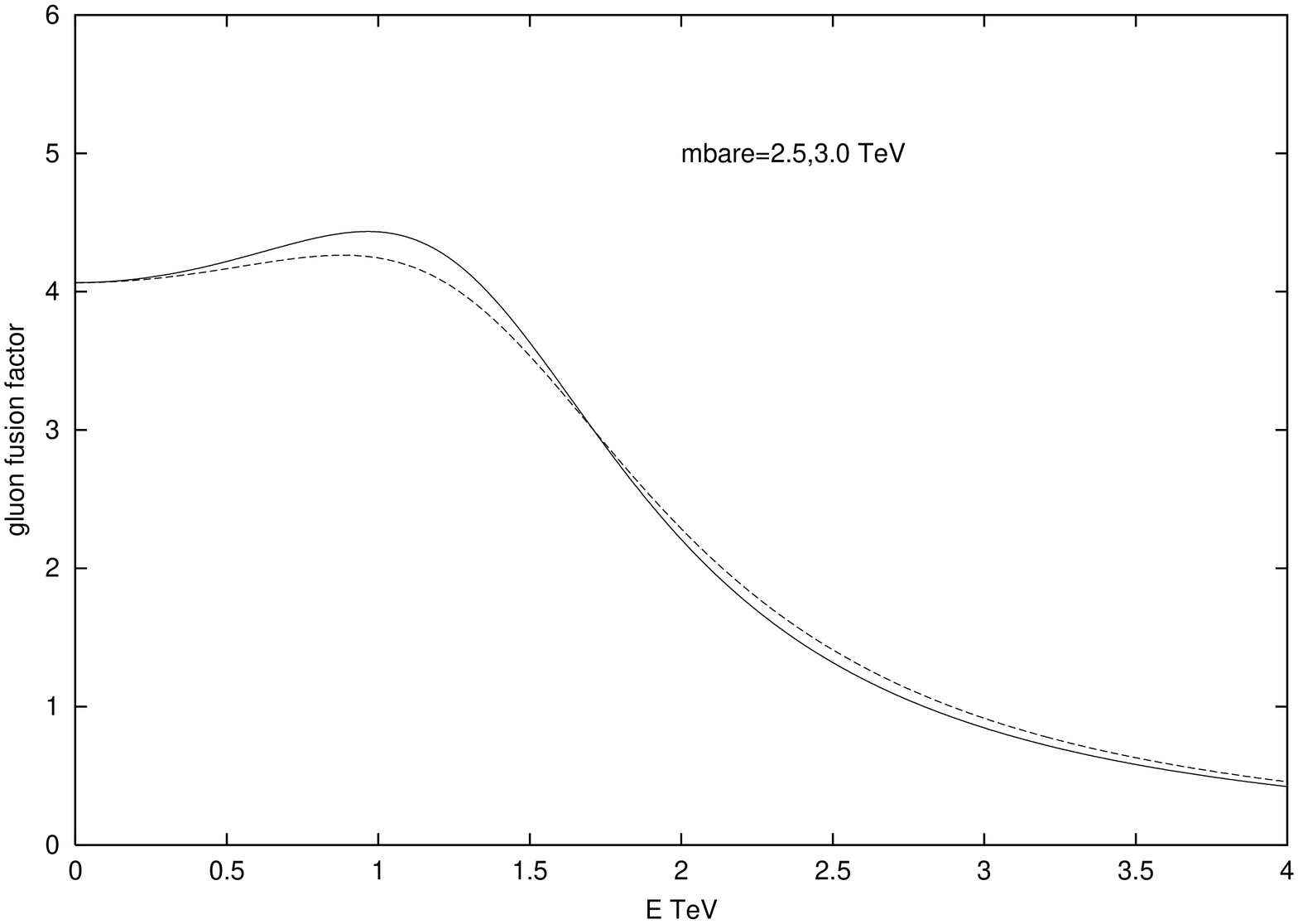,height=3in,width=3in, angle=270}} 
\end{center}
\caption
{Unitarized electroweak factor $\frac{  g_{\sigma \pi \pi} \cos
\delta_0^0 }{m_{\sigma b}^2 - s} $, plotted as a function of $E = \sqrt{s}$ for the cases
$m_{\sigma b} = 0.5, \, 1.0, \, 1.5, \, 2.0, \, 2.5, \, 3.0$ TeV.  }
\label{ewfactor_fig}
\end{figure}

\begin{figure}[htbp]
\vspace{0.5cm}
\begin{center}
\mbox{\epsfig{file=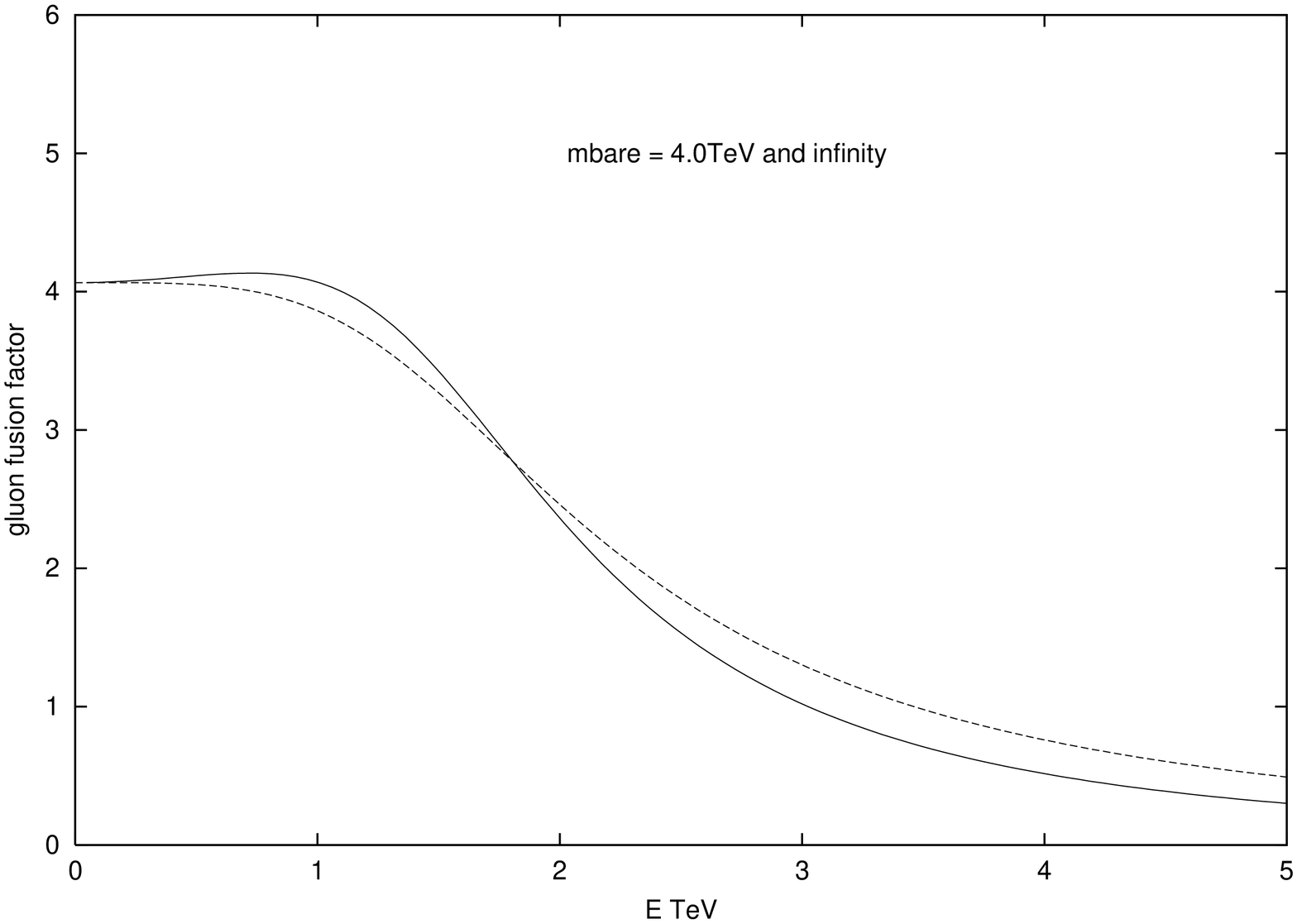,
height=3in,width=3in, angle=270}}
\end{center}
\caption
{Unitarized electroweak factor $\frac{  g_{\sigma \pi \pi} \cos
\delta_0^0 }{m_{\sigma b}^2 - s} $, plotted as a
 function of $E = \sqrt{s}$ for the cases
$m_{\sigma b} = 4,\infty$.  }
\label{ewfactorinfty_fig}
\end{figure}

It is especially interesting to examine the quantity ${\rm \cos} \delta_0^0
(s)$ which corresponds to the reduction in magnitude of the unitarized
gluon fusion amplitude from the one shown in Fig. \ref{glufu}.  Using
Eq. (\ref{pipiampl}) it is straightforward to find 
\begin{equation}
\cos \delta_0^0(s) = \frac { m_{\sigma b}^2 - s}{ { [ {\left
( m_{\sigma b}^2 - s \right)}^2 + {\left( \alpha(s)\left
( m_{\sigma b}^2 - s \right) + \beta \right)}^2 ]}^{\frac{1}{2}}}.
\label{cosdelta}
\end{equation} 
The numerator cancels the pole at $s = m_{\sigma b}^2$ in
 Eq. (\ref{FSI}) so one has the finite
result: 
\begin{equation}
[\frac{  g_{\sigma \pi \pi} }{m_{\sigma b}^2 - s} \cos \delta_0^0]
(s=m_{\sigma b}^2) = \frac{32 \pi v}{3 m_\sigma b}^2.
\label{treepole}
\end{equation}
It is clear that the factor ${\rm cos}\delta_0^0(s)$
 in Eq. (\ref{FSI}) effectively replaces the tree level denominator 
$(m_{\sigma b}^2-s)$ by the magnitude of the quantity in
Eq. (\ref{poleeq}), which defines the physical pole position
in the complex s plane and
 which we used to identify the physical Higgs mass. 
  Note that, as in the W-fusion situation, the regularization is {\it
required} for any value of $m_{\sigma b}$ (not just the large values)
in order to give physical meaning to the divergent expression.  The
regularized electroweak factor for the amplitude of Fig. \ref{glufu},
$\frac{  g_{\sigma \pi \pi} \cos \delta_0^0 }{m_{\sigma b}^2 - s} $,
is plotted as a function of $E = \sqrt{s}$
 in Fig. \ref{ewfactor_fig} for the cases
$m_{\sigma b} = 0.5, \, 1.0, \, 1.5, \, 2.0, \, 2.5, \, 3.0$ TeV.  It is
very interesting to observe that these graphs clearly
 show a peaking which is
correlated with the physical Higgs mass [taken, say, as $\sqrt{ {\rm
Re}(z_0)}$] rather than the bare Higgs mass, $m_{\sigma b}$.  At
$m_{\sigma b} = 0.5$ TeV one is still in the region where the physical
mass is close to $m_{\sigma b}$.  Already at $m_{\sigma b} = 1.0 $ TeV
the peak has markedly broadened and is located at about 0.89 TeV.  At
$m_{\sigma b} = 2.0$ TeV the still broader peak is located
 near 1.03 TeV, which is about as 
large as the physical mass ever gets.  Beyond
this, the peak continues to broaden but the location of the physical
mass goes to smaller $s$.  For example, at $m_{\sigma b} = 3.0$ TeV,
the peak is down to 0.88 TeV and is much less pronounced.  Going
further the curves do not look very different from the one with
$m_{\sigma b} = \infty$, shown in Fig. \ref{ewfactorinfty_fig}.
  Here the peaking has
disappeared and we have the analytic form (recall that $g_{\sigma
\pi \pi} = \frac{m_{\sigma b}^2}{v}$):
\begin{equation}
\lim_{m_{\sigma b} \rightarrow \infty} ( \frac {g_{\sigma \pi \pi}
\cos \delta_0^0 (s)}{m_{\sigma b}^2 - s} ) = { [ v^2 + {( \frac{s}{16
\pi v} )}^2 ]}^{-\frac{1}{2}}.
\end{equation}
This is the analog of Eq. (\ref{unlsm}) and corresponds to the
unitarized minimal non-linear sigma model. It still
 takes into account final state interaction effects in direct production 
of WW or ZZ pairs from gg fusion. The non-trivial structure
is seen to be confined to the region $s^{1/2}$ less than about 5 TeV.  
There might be a sense in which such a prescription
 could be appropriate for
a situation with an arbitrarily heavy Higgs boson \cite{BDV91}.

\begin{figure}
\vspace{0.5cm}
\begin{center}
\mbox{\epsfig{file=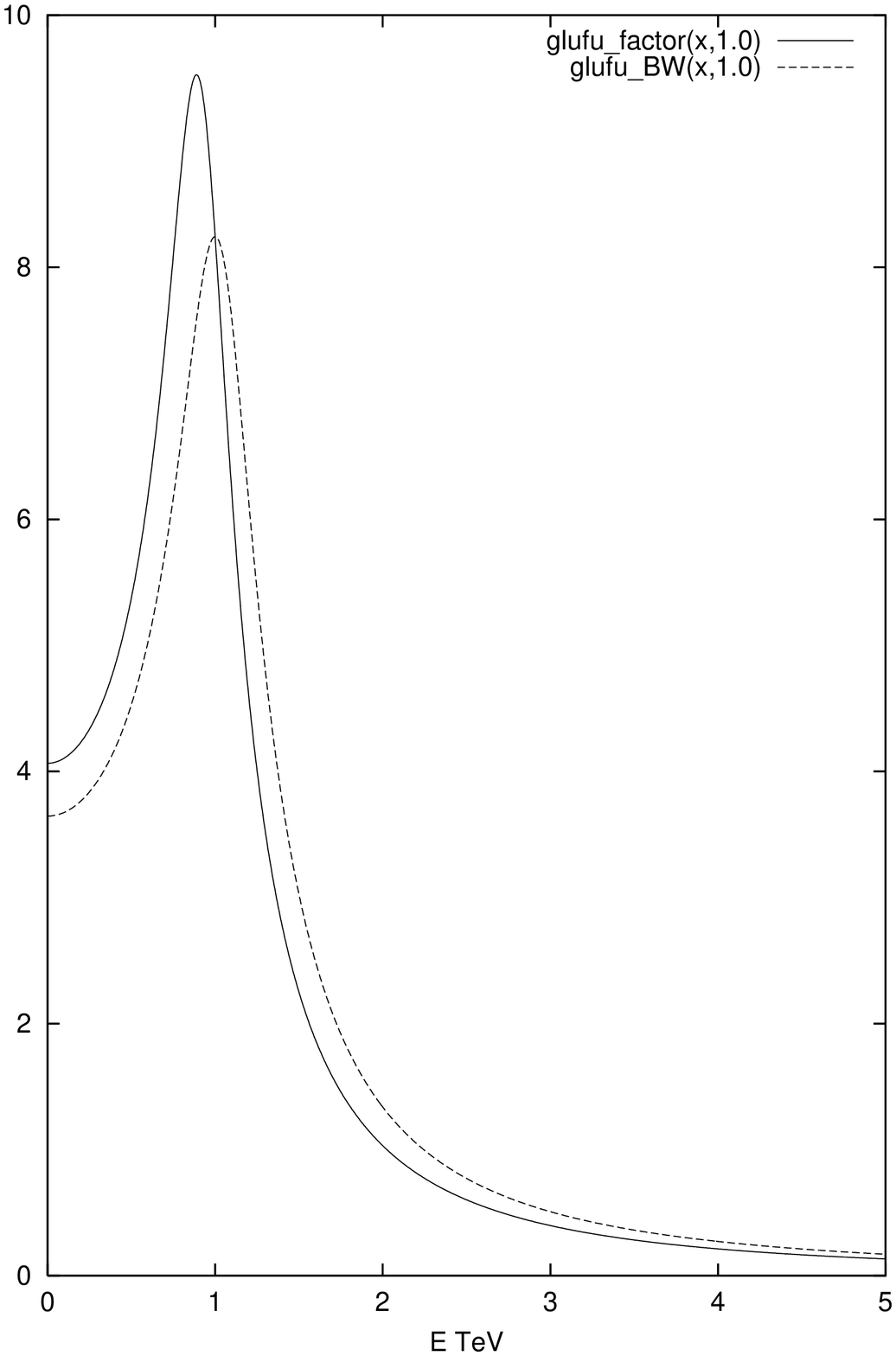,height=3in,width=3in, angle=0}}
\mbox{\epsfig{file=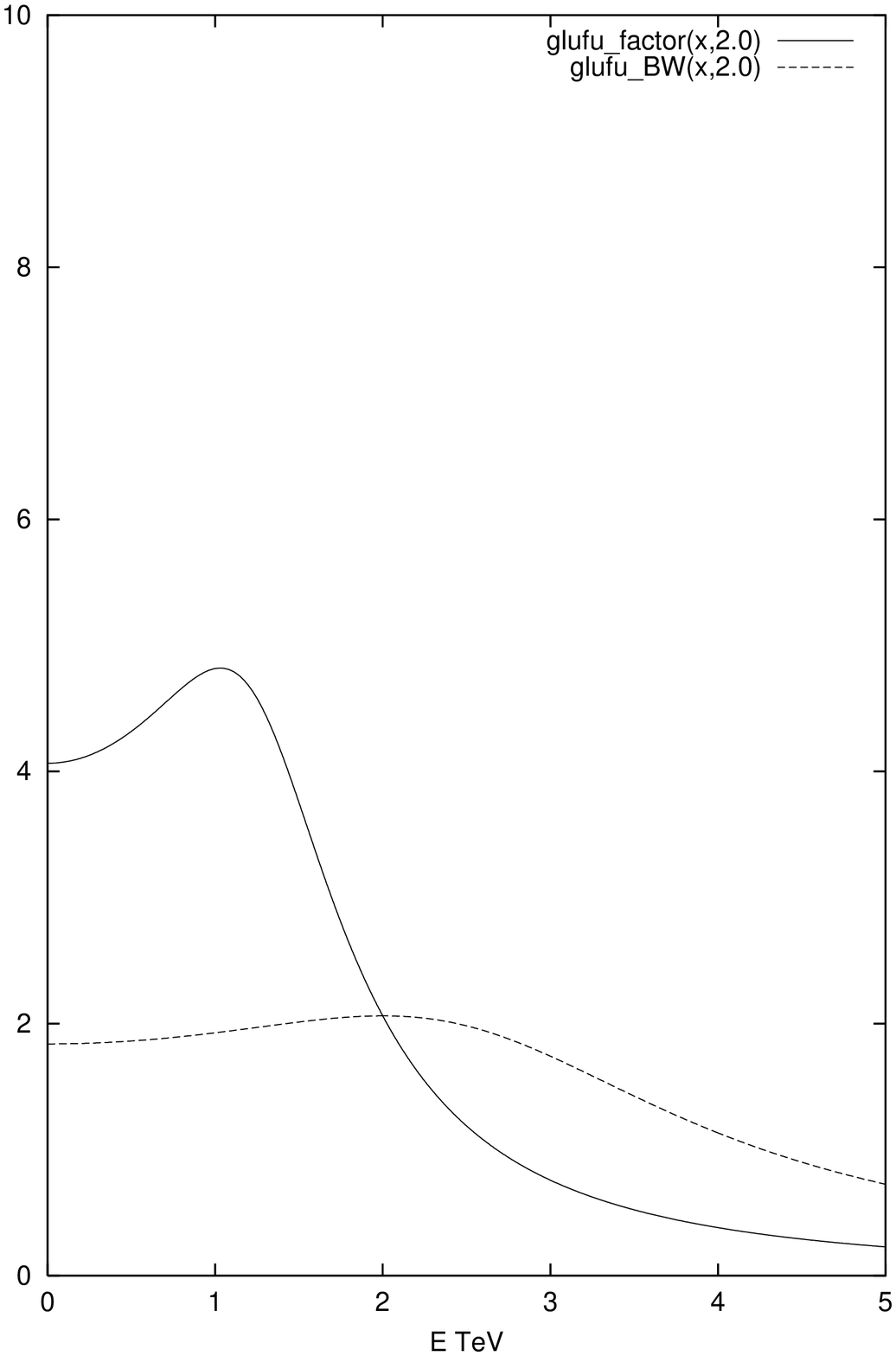,height=3in,width=3in, angle=0}}  
\mbox{\epsfig{file=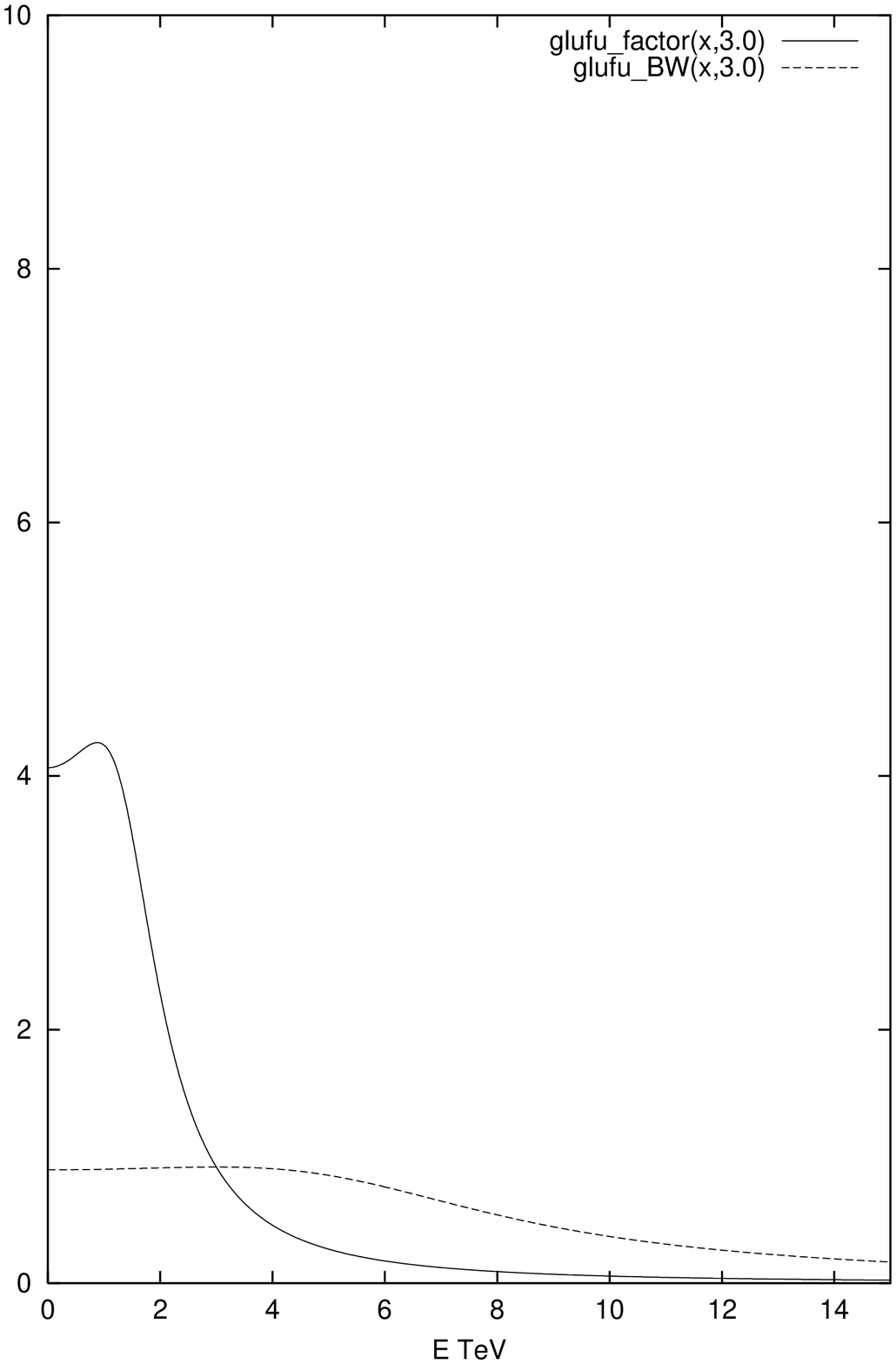,height=3in,width=3in, angle=0}}  
\end{center}
\caption
{Comparison of modulus of unitarized electroweak factors in Higgs
production amplitude: (i)  $\frac{g_{\sigma \pi
\pi}\cos{\delta_0^0}}{m_{\sigma b}^2 - s}$ for the present regularization
scheme 
(solid)  and (ii) prescription of including the tree-level
Higgs width (dashed) shown in Eq. (\ref{conventionalreg}).  Here
the three graphs correspond to  
$m_{\sigma b} =
1,2,3$ TeV.}
\label{glufu-BW}
\end{figure}

It is amusing to compare the effect of the present regularization
prescription, Eq. (\ref{FSI}) for the divergence at $s=m_{\sigma b}^2$
with that of the conventional Breit Wigner prescription used for
example in \cite{GvdB89,KD91}, given
in Eq. (\ref{conventionalreg}) with $\Gamma = \Gamma_{tree}$. These
are plotted for $m_{\sigma b} =1,2,3$ TeV in Fig. \ref{glufu-BW}.
The main observation is that, although the absolute value of
the factor in Eq. (\ref{conventionalreg}) flattens out due to the fact
that the tree-level Higgs width (to vector bosons) increases cubically
so the Higgs signal gets lost for larger $m_{\sigma b}$, the factor in
Eq. (\ref{FSI}) still has a peak at lower energies. Also 
 the magnitude of the modified factor suggested in Eq. (\ref{FSI}) is
larger than the corresponding Breit Wigner factor.  Other alternatives
to the Breit-Wigner prescription for treating the divergence at $s =
m_{\sigma b}^2$ in the gluon fusion Higgs production mechanism in
Fig. \ref{glufu} are momentum-dependent modifications of the
Higgs width \cite{VW92,S95} and explicit calculation of radiative 
corrections \cite{GvdB96}. 

\begin{figure}[htbp]
\vspace{0.5cm}
\begin{center}
\mbox{\epsfig{file=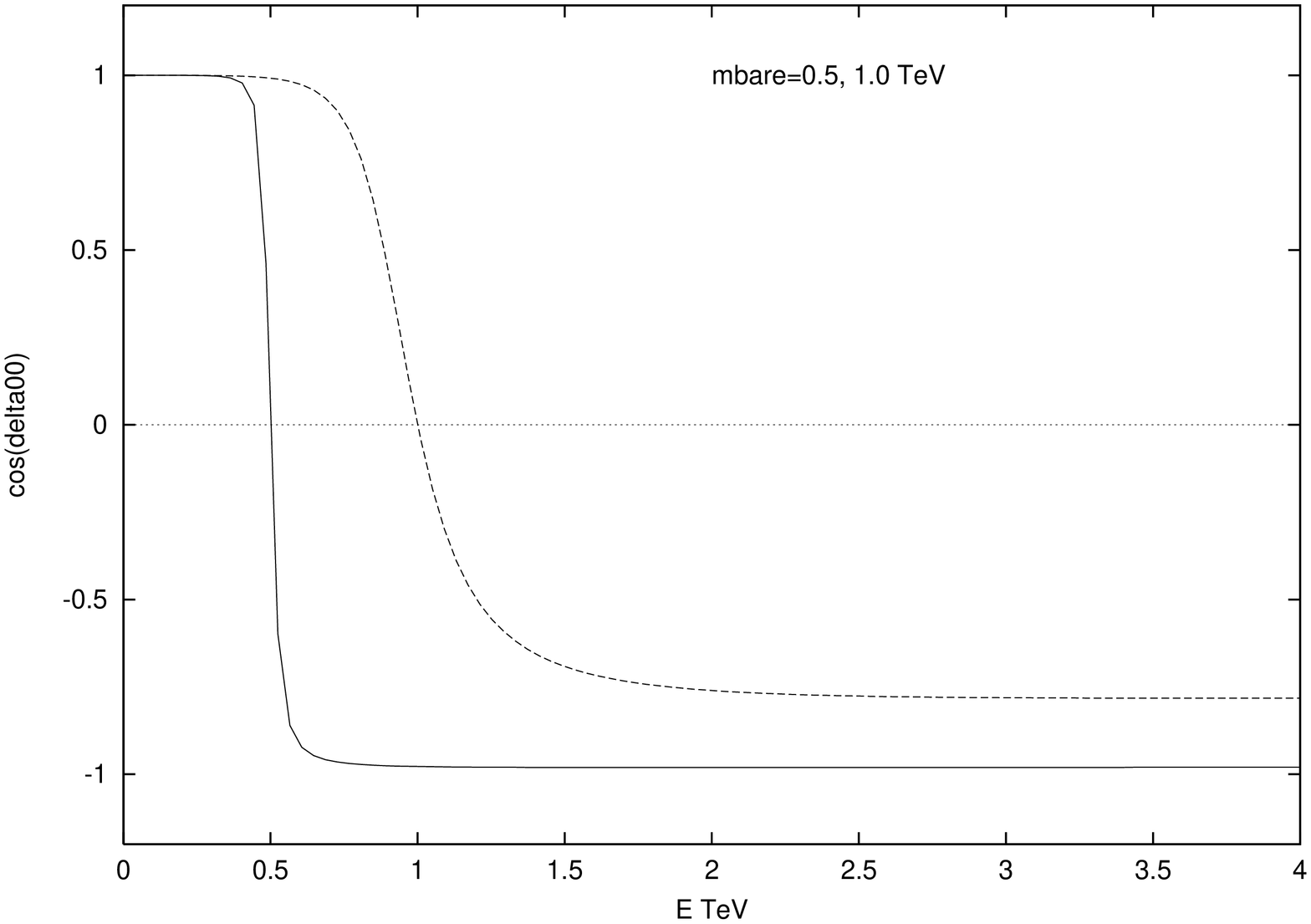,height=3in,width=3in, angle=270}}
\mbox{\epsfig{file=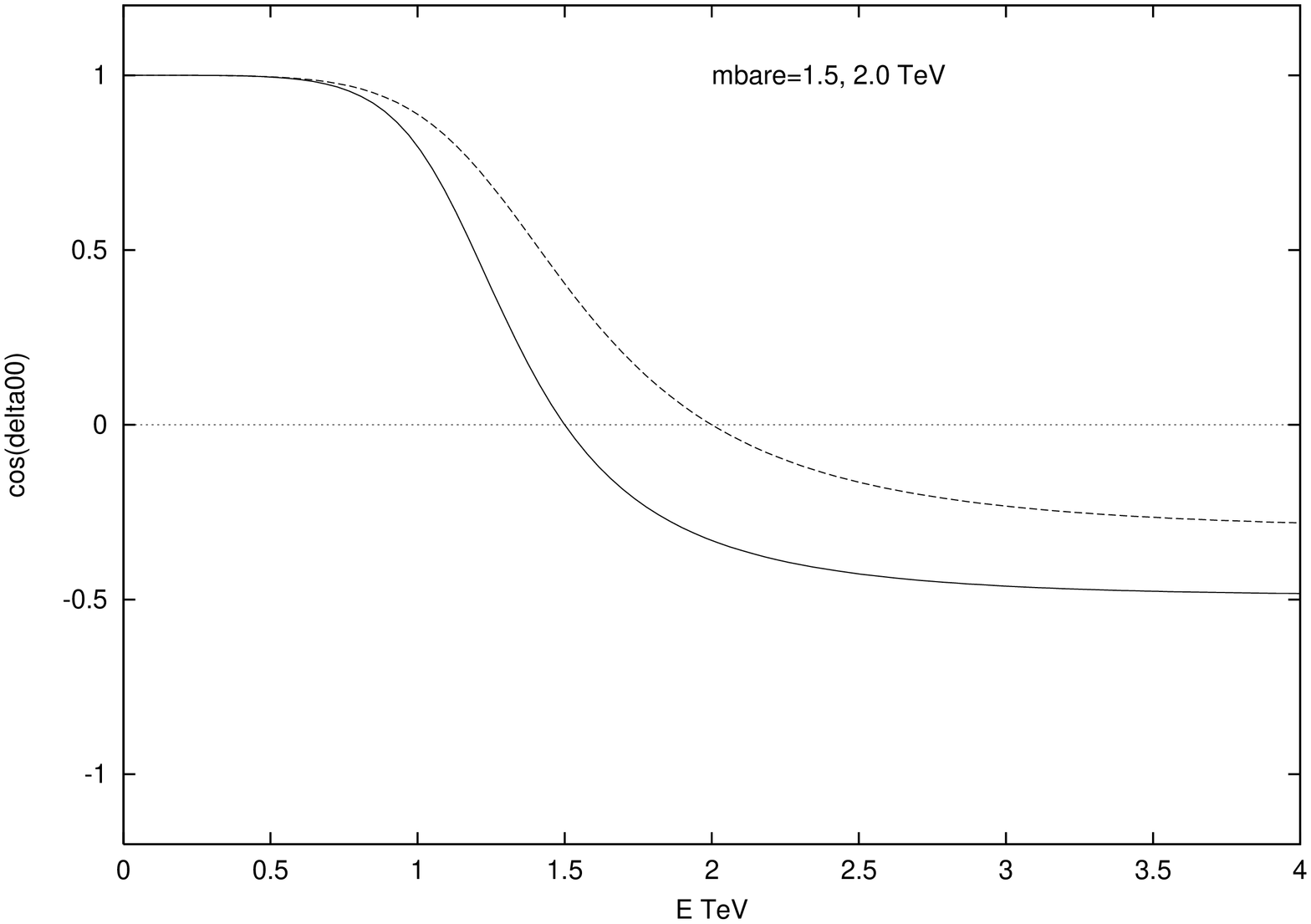,height=3in,width=3in, angle=270}}
\mbox{\epsfig{file=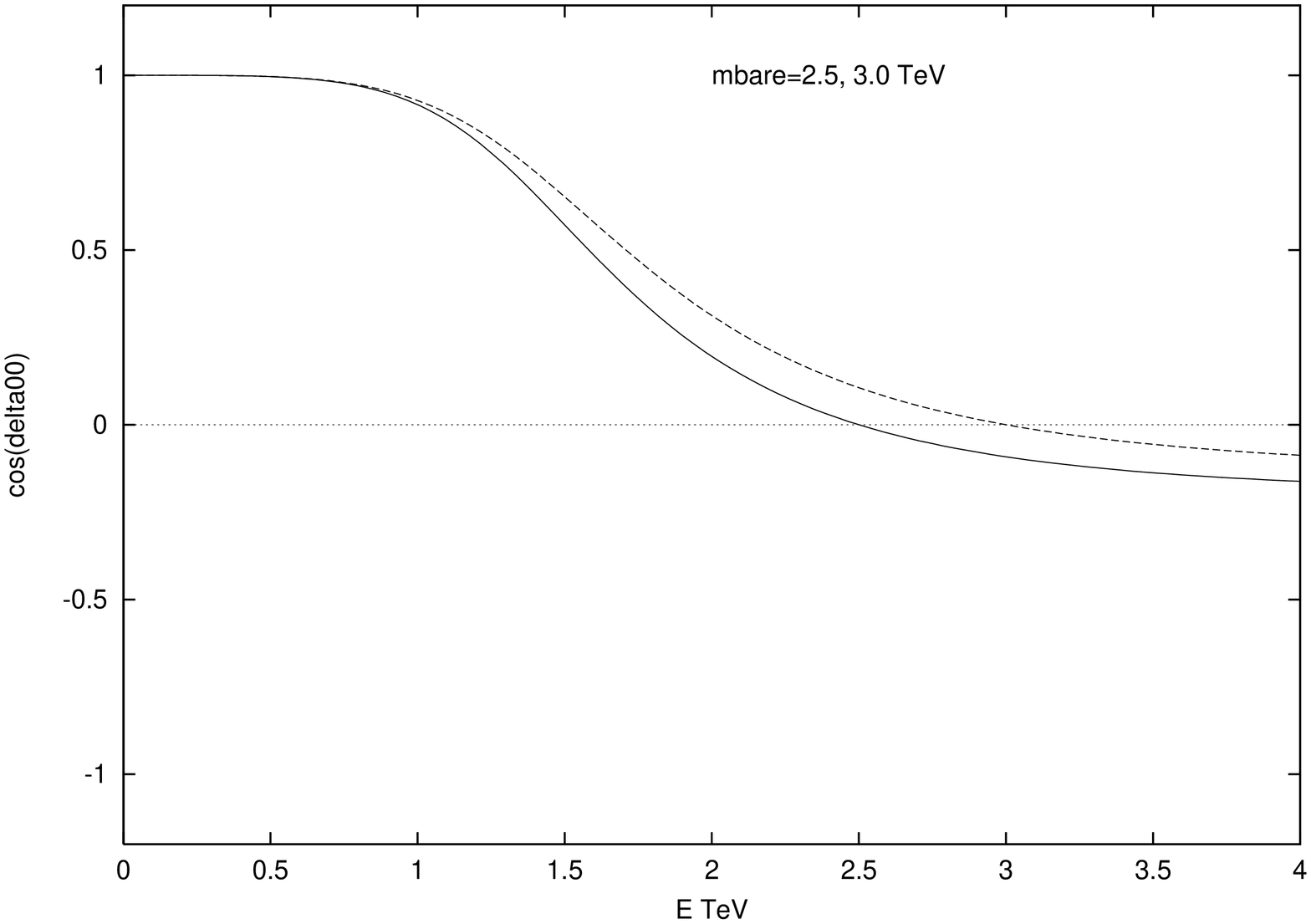,height=3in,width=3in, angle=270}}  
\end{center}
\caption
{Plots of $cos(\delta_0^0(s))$ for $m_{\sigma 
b}=0.5,1.0,1.5,2.0,2.5,3.0$TeV }
\label{delta00_fig}
\end{figure}

Of course, the electroweak amplitude factor we have been discussing
must be folded together with the triangle and gluon part of
Fig. \ref{glufu} as well as the gluon wavefunctions of the initiating
particles.  Furthermore only the I=J=0 partial wave amplitude has been
considered.  If various partial wave amplitudes are added, the ``final
state interaction'' phase factor in Eq. (\ref{FSI}), $\exp {[i
\delta_0^0 (s)]}$ must be included too.  The value of $\delta_0^0(s)$
may be readily obtained from Eq. (\ref{cosdelta}). ${\rm cos}\delta_0^0(s)$
is plotted in Fig. \ref{delta00_fig} for representative
values of  $m_{\sigma b}$. For a 
detailed
practical implementation of this model it would be
appropriate to take into account the relatively strong coupling of 
the Higgs boson to the $t{\bar t}$ channel. This could be conveniently 
accomplished by unitarizing the two channel $W_LW_L-t{\bar t}$
scattering matrix.

On noting that, as we have illustrated (see also
Fig. \ref{PhysicalHiggsMass}), there are two bare mass values (with
different widths) for each physical mass one might wonder
if a very light Higgs boson could exist in the strongly interacting
mode.  Possibly its contribution to ``precision electroweak
corrections'' would be comparable to those of their light bare mass
images.  This seems interesting, although the corresponding bare
masses would be very large ($> 10$ TeV according to
Fig. \ref{PhysicalHiggsMass}), beyond where the validity of the model 
has been tested by the QCD analog. For bare masses in this
region it would seem most reasonable to approximate the situation
by using the $m_{\sigma b} \rightarrow \infty$ case.

In this paper we first noted that the K-matrix unitarized linear
SU(2) sigma model could explain the experimental
data in the scalar channel of QCD up to about 800 MeV. Since it is
just a scaled version of the minimal electroweak Higgs sector, which is
often treated with the same unitarization method, we concluded that there
is support for this approach in the electroweak model up to at least
Higgs bare mass  about 2 TeV. We noted that the
 relevant QCD effective Lagrangian needed to go higher in energy
is more complicated than the SU(2) linear sigma model and is
better approximated by the linear SU(3) sigma model. This
enabled us to extend the energy range of experimental
agreement at the QCD level
by including another scalar resonance. Similarly in the
 electroweak theory
there are many candidates - e.g. larger Higgs sectors, larger
gauge groups, supersymmetry, grand unified theories, technicolor,
string models and recently, symmetry breaking by background
chemical potentials \cite{mu} - which may give rise to more than one 
Higgs particle in the same channel. We interpreted the better
agreement at larger energies in the QCD model as also giving support
to a similar treatment for a perhaps (to be seen in the future)
 more realistic Higgs sector in the electroweak theory which
may be valid at higher energies due to
additional higher mass resonances.
Nevertheless we noted that even with one resonance, the minimal
K matrix unitarized model behaved smoothly at large bare mass by
effectively "integrating out" the Higgs while preserving unitarity. 

With added confidence in this simple approach we made a survey of
the Higgs sector for the full range of bare Higgs mass. While a lot 
 of work in this area has been done in the past we believe that some new 
points were added. In particular, we have noted that in this scheme the
characteristic factor of the W-W fusion mechanism for
Higgs production peaks at the bare mass of the Higgs boson,
while the characteristic factor for the gluon fusion mechanism
peaks at the generally lower {\it physical} mass.

It should be remarked that while the simplest K-matrix method of unitarization 
used here seems to work reasonably well, it is not
at all unique.  For example, Eq. (\ref{Tregularization}) might be
replaced by 
\begin{equation}
S_0^0(s) = \frac{ 1 + i ( {[T_0^0]}_{\rm tree} + f(s) )    }  {1 - i
( {[T_0^0]}_{\rm tree} + f(s) )}, 
\label{generalizedKmatrix}
\end{equation}
where $f(s)$ is an arbitrary real function.  However the fit in Section III
is good up to scaled energy, $ {\bar s} = \frac {s}{\sqrt {2} v} \approx 6$.
Thus $f(s)$ is expected to be small, at least for this energy range.
A simple modification \cite{DR} in this framework is to take
$f(s)=$ Re$[T^0_0(s)]_{1 loop}$. Of course this is not guaranteed to be more
accurate at large s in the non perturbative region. Other approaches
include using the large N approximation \cite{W91}, the Pade
approximant 
method \cite{DHT90,DR,W91}, the Inverse Amplitude method \cite{DHT90,IAM},
variational approaches \cite{S01} and the N/D method \cite{O99}.
It is beyond the scope of this paper to compare the different
approaches.

\acknowledgments
 We are happy to thank Masayasu Harada and Francesco
Sannino for very helpful discussions. 
The work of A. A-R. S. N. and J. S. has been supported in part by the US DOE
under contract DE-FG-02-85ER 40231.  D. B. wishes to acknowledge
support from the Thomas Jefferson National Accelerator Facility
operated by the Southeastern Universities Research Association (SURA)
under DOE Contract No. DE-AC05-84ER40150.
The work of A. H. F. has been supported by grants from the State
of New York/UUP Professional Development Committee, and the 2002
Faculty Grant from the School of Arts and Sciences, SUNY Institute
of Technology.  S.N  is supported in part by an ITP Graduate Fellowship.


\begin{thebibliography}{10}

\bibitem{kyotoconf}See the dedicated conference proceedings, S. Ishida et al
``Possible existence of the sigma meson and its implication to hadron
physics", KEK Proceedings 2000-4, Soryyushiron Kenkyu 102, No. 5, 2001.
Additional points of view are expressed in the proceedings, D. Amelin
and A.M. Zaitsev ``Hadron Spectroscopy'', Ninth International Conference
on Hadron Spectroscopy, Protvino, Russia(2001).

\bibitem{vanBev} E. van Beveren, T.A. Rijken, K. Metzger,
C. Dullemond, G. Rupp and J.E. Ribeiro, Z. Phys. {\bf C30}, 615
(1986). E. van Beveren and G. Rupp, hep-ph/9806246, 248.  See also
J.J. de Swart, P.M.M. Maessen and T.A. Rijken, U.S./Japan Seminar on the
YN Interaction, Maui, 1993 [Nijmegen report THEF-NYM 9403].

\bibitem{MP}
D. Morgan and M. Pennington, Phys. Rev. {\bf D48},  1185  (1993).

\bibitem{BMPV} A.A. Bolokhov, A.N. Manashov, M.V. Polyakov and
V.V. Vereshagin, Phys. Rev. {\bf D48}, 3090 (1993).  See also
V.A. Andrianov and A.N. Manashov, Mod. Phys. Lett. {\bf A8}, 2199
(1993).  Extension of this string-like approach to the $\pi K$ case
has been made in V.V. Vereshagin, Phys. Rev. {\bf D55}, 5349 (1997)
and  in A.V. Vereshagin and V.V. Vereshagin, {\it{ibid.}} {\bf
59}, 016002 (1999). 

\bibitem{AS94} N.N. Achasov and G.N. Shestakov, Phys. Rev. {\bf
D49}, 5779 (1994). 

\bibitem{Kam94}{R. Kam\'inski}, {L. Le\'sniak} and J. P. Maillet,  
Phys. Rev. {\bf D50}, 3145 (1994).

\bibitem{SS}F.~Sannino and J.~Schechter, Phys. Rev.  {\bf D52},  96  (1995).

\bibitem{T}{N.A.~T\"ornqvist}, Z. Phys.
{\bf C68}, 647 (1995) and references therein.  In addition see
{N.A.~T\"ornqvist} and M. Roos, Phys. Rev. Lett. {\bf 76}, 1575
(1996), N.A. T\"ornqvist, hep-ph/9711483 and Phys. Lett. {\bf B426} 105 (1998).

\bibitem{DS} R. Delbourgo and M.D. Scadron, Mod. Phys. Lett. {\bf
A10}, 251 (1995).  See also D. Atkinson, M. Harada and A.I. Sanda,
Phys.~Rev. {\bf D46}, 3884 (1992).

\bibitem{JPHS}
{G.~Janssen, B.C.~Pearce, K.~Holinde and J.~Speth}, Phys. Rev. {\bf
D52},  2690  (1995).

\bibitem{Sv} M. Svec, Phys. Rev. {\bf D53}, 2343 (1996).

\bibitem{Ishida} S. Ishida, M.Y. Ishida, H. Takahashi, T. Ishida,
K. Takamatsu and T Tsuru, Prog. Theor. Phys. {\bf 95}, 745 (1996),  
S.~Ishida, M.~Ishida, T.~Ishida, K.~Takamatsu and T.~Tsuru,
Prog. Theor. Phys. {\bf 98}, 621 (1997). See also M. Ishida and S. Ishida,
Talk given at 7th International Conference on Hadron Spectroscopy (Hadron
97), Upton, NY, 25-30 Aug. 1997, hep-ph/9712231.

\bibitem{HSS1}M. Harada, F. Sannino and J. Schechter, Phys. Rev. {\bf D54},
1991 (1996).

\bibitem{HSS2}M. Harada, F. Sannino and J. Schechter,
Phys. Rev. Lett. {\bf 78}, 1603 (1997).

\bibitem{BFSS1}D. Black, A.H. Fariborz, F. Sannino and J. Schechter,
Phys. Rev. {\bf D58}, 054012 (1998).

\bibitem{BFSS2}D. Black, A.H. Fariborz, F. Sannino and J. Schechter,
 Phys. Rev. {\bf D59}, 074026 (1999).

\bibitem{OOP} J.A. Oller, E. Oset and J.R. Pelaez, Phys. Rev. Lett.
{\bf 80}, 3452 (1998). See also K. Igi and K. Hikasa, Phys. Rev. {\bf
D59}, 034005 (1999).

\bibitem{AnSa}A.V. Anisovich and A.V. Sarantsev, Phys. Lett. {\bf B413},
137 (1997).

\bibitem{EFSS}V. Elias, A.H. Fariborz, Fang Shi and T.G. Steele, 
Nucl. Phys. {\bf A633}, 279 (1998).

\bibitem{Dm} V. Dmitrasinovi\'c, Phys. Rev. {\bf C53}, 1383 (1996).

\bibitem{MO}P. Minkowski and W. Ochs, Eur. Phys. J. {\bf C9}, 283 (1999).

\bibitem{GN}S. Godfrey and J. Napolitano, hep-ph/9811410.

\bibitem{BG}L. Burakovsky and T. Goldman, Phys. Rev. {\bf D57}
2879 (1998)

\bibitem{FS1}A. H. Fariborz and J. Schechter, Phys. Rev
{\bf D60}, 034002 (1999).

\bibitem{BFS2}D. Black, A. H. Fariborz and J. Schechter,
Phys. Rev. {\bf D61} 074030 (2000).
See also V. Bernard, N. Kaiser and U-G. Meissner,
Phys. Rev. D {\bf 44}, 3698 (1991). 

\bibitem{BFS3}D. Black, A. H. Fariborz and J. Schechter, 
Phys. Rev. {\bf D61} 074001 (2000).

\bibitem{Shakin}L. Celenza, S-f Gao, B. Huang and C.M. Shakin,
Phys. Rev. C {\rm 61}, 035201 (2000).


\bibitem{BFMNS01}D. Black,
A.H. Fariborz, S. Moussa, S. Nasri and J. Schechter,
 Phys.Rev. {\bf D64}, 014031 (2001).

\bibitem{Jaffe}R. L. Jaffe, Phys. Rev. {\bf D15}, 367 (1977).

\bibitem{mixing}In addition to \cite{BFS3} and \cite{BFMNS01}
above see T. Teshima, I. Kitamura and N. Morisita, J. Phys. G.
{\bf 28}, L391 (2002); F. Close and N. Tornqvist, {\it ibid.}
{\bf 28}, R249 (2002) and A. H. Fariborz, hep-ph/0302133.

\bibitem{sectionV}See section V of \cite{BFMNS01} above.

\bibitem{GL}M. Gell-Mann and M. L\'evy, Nuovo Cimento {\bf 16}, 705 (1960).

\bibitem{nonlinear} In addition to \cite{GL} above see
J. Cronin, Phys. 
Rev. {\bf 161}, 1483 (1967);
S. Weinberg, Phys. Rev. Lett {\bf 18}, 188 (1967).

\bibitem{chpt}J. Gasser and H. Leutwyler, Ann. Phy. {\bf 158},
142 (1984).

\bibitem{loops} A. Abdel-Rehim, D. Black, A. H. Fariborz
and J. Schechter, hep-ph/0304177.

\bibitem{CLT74}J.M. Cornwall, D.N. Levin and G. Tiktopoulos,
Phys. Rev. {\bf D10}, 1145 (1974).

\bibitem{LQT77}B.W. Lee, C. Quigg and H.B. Thacker, Phys. Rev. {\bf
D16} 1519 (1977).

\bibitem{CG85}M.S. Chanowitz and M.K. Gaillard, Nucl. Phys. {\bf B261}
(1985) 379.

\bibitem{BS90}J. Bagger and C. Schmidt, Phys. Rev. {\bf D 41}, 264 (1990).

\bibitem{CDP02}S.De Curtis, D. Dominici and J.R. Pelaez, hep-ph/0211353.

\bibitem{RS89}W. W. Repko and C. S. Suchyta, III, Phys. Rev. Lett.
{\bf 62}, 859 (1989).

\bibitem{DR}D. A. Dicus and W. W. Repko Phys. Lett. B {\bf 228},
503 (1989); Phys. Rev. {\bf D42}, 3660 (1990).

\bibitem{W91}S. Willenbrock, Phys. Rev. {\bf D43} 1710 (1991).

\bibitem{C02}M. S. Chanowitz, Phys. Rev. {\bf D66} 073002 (2002) and hep-ph/0304199.

\bibitem{LEP}The LEP Collaborations, the LEP Electroweak Working Group
and the SLD Heavy Flavor and Electroweak Groups, hep-ex/0212036.

\bibitem{NuT}G. P. Zeller et al, Phys. Rev. Lett. {\bf 88}, 091802
 (2002).

\bibitem{LOTW}W. Loinaz, N. Okamura, T. Takeuchi and L. C. R. 
Wijewardhana, Phys. Rev. {\bf D67} 073012 (2003).

\bibitem{DGMP97}For a discussion of how gauge boson scattering may be
incorporated into the $p \bar p$ annihilation cross section see for
example pages 197-199 of {\it Effective Lagrangians for the Standard
Model}, A. Dobado, A. G\'omez-Nicola, A.L. Maroto and J.R. Pel\'aez,
Springer-Verlag Berlin Heidelberg 1997.

\bibitem{GGMN78}H.M. Georgi, S.L. Glashow, M.E. Machacek and  
D.V. Nanopoulos, Phys. Rev. Lett. {\bf 40}, 692 (1978).

\bibitem{Chung}S. U. Chung {\it et al}, Ann. Physik {\bf 4} 404 (1995).



\bibitem{pipidata}E.A. Alekseeva {\it et al}., Sov. Phys. JETP {\bf
55}, 591 (1982), G. Grayer {\it et al}., Nucl. Phys. {\bf B75}, 189 (1974).

\bibitem{Levy}M. L\'evy, Nuovo Cimento {\bf 52A}, 23 (1967).
  See S. Gasiorowicz and D. A. Geffen, Rev. Mod. Phys.
 {\bf 41}, 531 (1969) for a review which contains a large bibliography.

\bibitem{SU1}J. Schechter and Y. Ueda, Phys. Rev. {\bf D3}, 2874,
1971; Erratum D {\bf 8} 987 (1973).  See also J. Schechter and
Y. Ueda, Phys. Rev. {\bf D3}, 168, (1971).

\bibitem{SU2}J. Schechter and Y. Ueda, Phys. Rev. D {\bf 4}, 733 (1971).

\bibitem{CH}See also L.H. Chan and R. W. Haymaker, Phys. Rev.
 {\bf 07}, 402 (1973); {\bf 10}, 4170 (1974).

\bibitem{leconstants}
J. Donoghue, C. Ramirez and G. Valencia, Phys.Rev. {\bf D39}
, 1947 (1989); G. Ecker, J. Gasser, A. Pich and E. de Rafael,
Nucl. Phys. {\bf B321}, 311 (1989); G. Ecker, J. Gasser,
H. Leutwyler, A. Pich and
E. de Rafael, Phys. Lett. {\bf B233}, 425 (1989).


\bibitem{H72}W. Hudnall, Phys. Rev. {\bf D6}, 1953 (1972).


\bibitem{DW89}For the Standard Model case it has been explicitly verified
that the one-loop corrections to the J=0 WW/ZZ scattering amplitude,
discussed in Section V, become large for $m_H > 1$ TeV.  See S. Dawson and S. Willenbrock, Phys. Rev. Lett. {\bf
62} 1232, 1989.


\bibitem{pionmass}
In the case where the pion mass
 is retained in Eqs. (\ref{alphabeta}), $-{\rm Im} z_0$ behaves
 identically as in the
$m_\pi = 0$ case while the ``physical'' sigma mass, $\sqrt {{\rm Re}  
z_0}$, approaches the small constant value 0.22 GeV.  The ``physical''
sigma width, $\frac{-Im z_0}{\sqrt{ Re z_0}}$ approaches the
relatively large constant value 1.96 GeV.  Qualitatively, this is in
agreement with the $m_\pi = 0$ case.

\bibitem{DH89}A. Dobado and M.J. Herrero, Phys. Lett {\bf B228} 495
(1989).

\bibitem{DoR90}J.F. Donoghue and C. Ramirez, Phys. Lett. {\bf B234},
361 (1990).

\bibitem{zerompi}This best fit value of $\bar m$ was found when the
physical pion mass is taken.  As discussed in Section IV, the QCD
results do not change too much if we set $m_\pi = 0$.

\bibitem{massdecrease}The fact that $\sqrt
{ {\rm Re}z_0} < m_{\sigma b}$ can be understood as due to the
opposite signs of $\alpha(s)$ and $\beta(s)$ in Eq. (\ref{pipiampl}).
  This is explained in more detail in the
next-to-last paragraph of Section III in \cite{BFMNS01}.

\bibitem{CDG84}R. Casalbuoni, D. Dominici and R. Gatto, Phys. Lett.
B {\bf 147}, 419 (1984).
                                              
\bibitem{Pa00}J. Pasupathy, Mod. Phys. Lett. A{\bf 12}, 1605 (2000).


\bibitem{GHPTW87}J.F. Gunion, H.E. Haber, F.E. Paige, Wu-Ki Tung and
S.S.D. Willenbrock, Nucl. Phys. {\bf B294}, 621 (1987).

\bibitem{D88} D.A. Dicus, Phys. Rev. {\bf D
38}, 394 (1988).

\bibitem{GvdB89}E.W.N.
Glover and J.J. van der Bij, Phys. Lett {\bf B219} 488 (1989).  E.W.N.
Glover and J.J. van der Bij, Nucl. Phys. {\bf B321} (1989) 561.

\bibitem{KD91}C. Kao and D.A. Dicus, Phys. Rev. {\bf D43} 1555 (1991).  

\bibitem{BDV91}The one-loop result for this case is given in J. Bagger, S. Dawson and G. Valencia,
Phys. Rev. Lett. {\bf 67}, 2256 (1991). 

\bibitem{VW92}G. Valencia and S. Willenbrock, Phys. Rev. {\bf D
46}, 2247 (1992).

\bibitem{S95}M.H. Seymour, Phys. Lett {\bf B 354} (1995) 409.

\bibitem{GvdB96}A. Ghinculov and J.J. van der Bij, Nucl. Phys. {\bf B
482} 59 (1996).  


\bibitem{mu}F. Sannino and K. Tuominen, hep-ph/0303167.
  
 
\bibitem{DHT90}A. Dobado, M.J. Herrero and T.N. Truong,
Phys. Lett. {\bf B 235}, 129 (1990), {\it ibid} {\bf B235, 134}
(1990).  A. Dobado {\it ibid} {\bf B 237}, 457 (1990).  A. Dobado,
M.J. Herrero and J. Terron, Zeit. Phys. {\bf C50}, 205 (1991) and 465
(1991).  

\bibitem{IAM} J.R. Pel\'aez, Phys. Rev. {\bf D55} 4193 (1997).


\bibitem{S01}F. Siringo, Europhys. Lett. 57 (2002) 820 - 826.



\bibitem{O99}J.A. Oller, Phys. Lett. {\bf B477} (2000) 187-194.


\end{thebibliography}
\end{document}